\documentclass[aps,
prl,
reprint,
showpacs,
superscriptaddress,
longbibliography,
 floatfix,
justify
]{revtex4-2}

\usepackage[utf8]{inputenc}
\usepackage[usenames,dvipsnames]{xcolor}
\usepackage{graphicx}
\usepackage{amsfonts}
\usepackage{amssymb}
\usepackage{blkarray}
\usepackage{amsmath}
\usepackage{mathtools}
\usepackage{bm}
\usepackage{dsfont}
\usepackage{physics}
\usepackage{braket}
\usepackage{tikz}
\usepackage{tikz-network}
\usepackage{tikz}
\usepackage{tikz-3dplot}
\usetikzlibrary{3d, calc}
\usetikzlibrary{decorations.markings}
\usetikzlibrary {arrows.meta}
\tikzset{
    edge/.style={line width=0.7pt, black},
    envedge/.style={line width=1.0pt, black},
    site/.style={circle, draw=black, fill=B, line width=0.7pt, minimum size=6pt, inner sep=0pt},
    envsite/.style={rectangle, draw=black, fill=gray, line width=0.7pt, minimum size=6pt, inner sep=0pt},
    lbl/.style={label distance=-0.1cm, font=\small},
    clbl/.style={label distance=-0.2cm, font=\small},
}

\tikzset{
    lattice3d/.style={
            scale=1.0,
            x={(-0.6cm, -0.60cm)},
            y={(1cm, 0cm)},
            z={(0cm, 1cm)},
        }
}

\tikzset{
    lattice3dv2/.style={
            scale=1.0,
            x={(-0.9cm, -0.5cm)},
            y={(1cm, 0cm)},
            z={(0cm, 1cm)},
        }
}
\tikzset{
    lattice3dv3/.style={
            scale=1.0,
            x={(-0.7cm, -0.5cm)},
            y={(1cm, 0cm)},
            z={(0cm, 1cm)},
        }
}
\usepackage{txfonts}
\usepackage{booktabs}
\usepackage{pifont}
\usepackage{ragged2e}
\usepackage{stmaryrd}
\usepackage[pdfusetitle,%
bookmarks=true,%
colorlinks,%
linkcolor=blue,%
citecolor=blue,%
urlcolor=blue%
]{hyperref}

\newcommand{\subfigref}[2]{\hyperref[fig:#1]{\ref*{fig:#1}(#2)}}

\newcommand{\TUM}{\affiliation{Technical University of Munich, TUM School of Natural Sciences, Physics Department, 85748 Garching, Germany}}
\newcommand{\MCQST}{\affiliation{Munich Center for Quantum Science and Technology (MCQST), Schellingstr. 4, 80799 M{\"u}nchen, Germany}}
\newcommand{\TCM}
{\affiliation{T.C.M. Group, Cavendish Laboratory, JJ Thomson Avenue, Cambridge CB3 0HE, UK}}
\newcommand{\UBC}{\affiliation{Department of Physics and Astronomy, and Quantum Matter Institute, University of
British Columbia, Vancouver, BC, Canada V6T 1Z1}}

\begin{document}

\def\papertitle{{Noise-resilient sequential circuits for generating quantum order}}
\title{\papertitle}
\author{Konstantin Gattinger} \thanks{These authors contributed equally to this work.\\} \TUM \MCQST 
\author{Julian Boesl} \thanks{These authors contributed equally to this work.\\} \TUM \MCQST 
\author{Max McGinley} \TCM \UBC
\author{Michael Knap} \TUM \MCQST
\author{Frank Pollmann}\TUM \MCQST
\date{\today}

\begin{abstract}
Sequential circuits provide an optimal linear-depth route to unitarily preparing long-range ordered quantum states, but circuit-level noise can be far more damaging than noise applied after preparation: local errors may be propagated by subsequent gates into nonlocal defects that destroy the target order.
In this work, we present classes of unitary sequential circuits that prepare certain non-trivial orders in the presence of Pauli noise. Our constructions exploit the spatial structure of the syndromes of the target state, which in certain cases allows us to systematically suppress the propagation of errors using strictly local gates. The strategy can be applied both to symmetry-breaking order, for which we present a 3D example, and to topological order, which we showcase on a 4D version of the toric code. We also highlight how this stability against errors necessitates non-Clifford gates, and how measurement-feedback loops can be used to stabilize lower-dimensional states as well.
\end{abstract}

\maketitle

\textbf{\emph{Introduction.---}} The challenge of preparing long-range entangled quantum states on digital quantum devices has attracted a lot of interest recently~\cite{satzinger_realizing_2021, bluvstein_quantum_2022, iqbal2024nonabelian, iqbal2025qutrit, acharya_quantum_2025, lo2026universalgatesbraidingfusing}. One approach is to use sequential circuits of local unitaries, whose depth scales with linear system size; this scaling is in fact optimal for purely unitary, local operations~\cite{bravyi_liebrobinson2006,yichen_quantum2015, chen_sequential2024}. In one dimension, all matrix product states can be prepared in this way~\cite{schon_sequential2005, schon_sequential2007, wei_generation2021, jones_skeleton2021, malz_preparation2024}, which has been used to study a quantum phase transition on a quantum device~\cite{smith_crossing2022}. Likewise, certain classes of higher-dimensional tensor network states guarantee the existence of such circuits~\cite{banuls_sequentially2008, zaletel_isometric_2020, slattery_quantum_2021, pichler2017universal, wei_sequential_2022}, including all string-net models, which serve as fixed points of many topological orders~\cite{levin_stringnet2005, soejima_isometric_2020, liu_methods2022, Chen2024quantumcircuits, liu_simulating2024, boesl_quantum2025, boesl2025skeleton}. Sequential circuits of this type have also been implemented experimentally to prepare topologically ordered states on quantum processors~~\cite{satzinger_realizing_2021, Xu2024FibonacciBraiding, Minev2025StringNetCondensation}.

However, the successful preparation of a non-trivial quantum state may easily be prevented by the presence of noise in a device. Local noise appearing during the sequential generation is, in fact, much more destructive than if it acted on an already prepared state: While the latter may be corrigible up to a certain threshold in the thermodynamic limit~\cite{shor_scheme1995, dennis2002topological, fowler_hightreshold2009, breuckmann2016localdecoders, fan_diagnostics_2024, xu_critical_2025, ellison_towards_2025, lavasani_stability2025, liu_detecting_2025, kim2026mixed}, in the former case the layers of unitaries can propagate the error. Those non-local perturbations in turn destroy the long-range entanglement at any finite noise strength. For instance, the 1D GHZ state can be prepared via a circuit of nearest-neighbor CNOT gates; however, its symmetry-breaking order is unstable to bit-flip noise~\cite{nielsen2010quantum}. Similarly, the circuit preparing the 2D toric code wavefunction is unstable to Pauli noise~\footnote{In Refs.~\cite{shukla_boson_2018, williamson_stability_2021}, it is shown that particular tensor network representations of toric code wavefunctions are unstable against noise on the virtual degrees of freedom; this can be reinterpreted as an instability of the sequential circuit.}. A relevant question is thus to find sequential circuits that can prepare these states in the presence of noise.

\begin{figure}[t]
    \centering
    \includegraphics[width=1\linewidth]{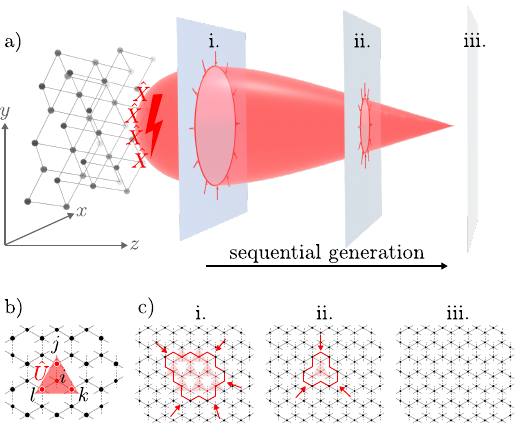}
    \caption{a) Cartoon picture of the circuit logic, exemplified in a 3D cubic system. Pauli $\hat{X}$ errors lead to closed membranes of syndromes (red surface). Locally, the sequential circuit has access to the curvature of the loop-like intersection with the layer and can shrink the membrane. b) The local gate $\hat{U}$ entangles the control qubits $\mathbf{j}, \mathbf{k},\mathbf{l}$ in layer $\mathbf{s}_j$ (large red points) with the unprepared qubit $\mathbf{i}$ in layer $\mathbf{s}_{j+1}$ (small red point). c) Illustration of three layers in the sequential preparation. Red sites indicate the erroneous domain, the red line being the associated domain wall. The circuit iteratively shrinks the domain according to a majority vote at the domain wall (i., ii.), cleaning it up after several steps (iii.).}
    \label{fig:fig1}
\end{figure}

In this work, we show that in sufficiently high spatial dimensions, it is possible to design simple quantum circuits for certain stabilizer states which are stable against Pauli noise. In these states, the syndromes related to the potentially pernicious errors appear on closed manifolds of dimension $d \geq 2$, allowing each layer of unitaries to shrink syndromes depending on their local curvature; see Fig.~\ref{fig:fig1} for a visualization. This class includes 3D symmetry-breaking order and a 4D version of the toric code, for which we provide explicit circuits and numerically confirm the stability of the non-trivial order in presence of noise. We also discuss why these stable circuits, in contrast to the most naive choices, require non-Clifford gates, and how stabilizer measurements and feedback extend the applicability of our approach to lower-dimensional states.

\begin{table}[t]
    \centering
    \begin{minipage}[t]{0.42\linewidth}
        \centering
        \textbf{a) SSB}\\[0.5ex]
        \begin{tabular}{ccc}
            \toprule
            $D$ & Sequential & Temperature \\
            \midrule
            1 & $\times$      & $\times$ \\
            2 & $\checkmark^*$    & \checkmark \\
            3 & \checkmark    & \checkmark \\
            \bottomrule
        \end{tabular}
    \end{minipage}
    \hfill
    \begin{minipage}[t]{0.52\linewidth}
        \centering
        \textbf{b) Toric code}\\[0.5ex]
        \begin{tabular}{ccc}
            \toprule
            $D$ & Sequential & Temperature \\
            \midrule
            2        & $\times$      & $\times$ \\
            3        & $\checkmark^*$    & $\times$ \\
            4 (2, 2) & $\checkmark^*$    & \checkmark \\
            4 (3, 1) & \checkmark    & $\times$ \\
            \bottomrule
        \end{tabular}
    \end{minipage}

    \caption{Stability of sequential generation against noise, and stability of the resulting state against thermal noise, in different spatial dimensions for a spontaneous symmetry-broken (SSB) state in a) and the topologically ordered toric code in b). A checkmark (\checkmark) indicates stability, while a cross ($\times$) indicates instability. For sequential generation, $\checkmark^*$ indicates stable sequential generation if stabilizer measurements and classical feedback are performed within each slice. In 4D, the toric code admits two distinct generalizations, one of which is thermally stable \cite{hastings2011topological}, and the other of which is stable under unitary sequential preparation.}
    \label{tab:stability}
\end{table}

\textbf{\emph{Noisy Sequential Quantum Circuits.---}} A large class of gapped quantum states can be generated by applying a sequential quantum circuit to a trivial input state $\ket{\Psi_\text{in}}$; such a circuit consists of local unitaries which act on each degree of freedom a finite number of times~\cite{satzinger_realizing_2021,liu_methods2022, chen_sequential2024}. If the targeted state $\ket{\Psi_\text{out}}$ is long-range entangled, the depth of such a preparation circuit scales at least with the linear system size $L$~\cite{bravyi_liebrobinson2006, yichen_quantum2015}. Different choices for the subregions each layer of unitaries acts on are possible. Here we consider a $D$ dimensional system of qubits which we split in $(D-1)$ dimensional slices $\mathbf{s}_j$. The initial state $\ket{\Psi_\text{in}} = \ket{\Psi_b}_{\mathbf{s}_1} \ket{0}_{\mathbf{s}_2 \cdots\mathbf{s}_L}$ is in the empty state except for the first slice $\mathbf{s}_1$ hosting a boundary state $\ket{\Psi_b}$. The circuit consists of sequentially applied unitaries $\hat{U}_{\mathbf{s}_{j}, \mathbf{s}_{j+1}}$ acting on two adjacent slices $\mathbf{s}_j, \mathbf{s}_{j+1}$, thereby entangling them. Each unitary $\hat{U}_{\mathbf{s}_{j}, \mathbf{s}_{j+1}}$ is further decomposed into local gates $\hat{U}^k_{\mathbf{s}_{j}, \mathbf{s}_{j+1}}$ which are applied in constant depth, $\hat{U}_{\mathbf{s}_{j}, \mathbf{s}_{j+1}} = \prod_k \hat{U}^k_{\mathbf{s}_{j}, \mathbf{s}_{j+1}}$. The unitary generation protocol is
\begin{equation}
 \vert \Psi_\mathrm{out}\rangle = \prod_{j=1}^{L-1} \hat{U}_{\mathbf{s}_{j}, \mathbf{s}_{j+1}} \vert \Psi_\mathrm{in} \rangle 
 \label{eq:SQCGeneral}
\end{equation}
with linear depth $\mathcal{O}(L)$; the unitaries are ordered such that operators with smaller $j$ are applied earlier. In passing we mention that this circuit architecture implies that the state $\ket{\Psi_\text{out}}$ can be expressed as an isometric tensor network state (isoTNS)~\cite{zaletel_isometric_2020, slattery_quantum_2021, wei_sequential_2022} (see Sec.~A of the supplemental material~\cite{SM}). Each of the unitaries $\hat{U}^k_{\mathbf{s}_{j}, \mathbf{s}_{j+1}}$ is further specified to take the qubits in its support on slice $\mathbf{s}_{j}$ exclusively as control qubits; i.e., a unitary is applied on the qubits in its support on slice $\mathbf{s}_{j+1}$ which is conditioned on the configurational basis state of $\mathbf{s}_{j}$.

In the presence of noise, errors will drive the generation protocol away from the clean output state $\ket{\Psi_\text{out}}$ and instead lead to a mixed state $\hat{\rho}_\text{out}$~\cite{fan_diagnostics_2024}. In this work, we will consider the case of Pauli noise acting on the qubits of slice $\mathbf{s}_j$ before the unitary $\hat{U}_{\mathbf{s}_{j}, \mathbf{s}_{j+1}}$ is applied, represented by a channel $\mathcal{E}_j = \bigotimes_{\mathbf{i} \in \mathbf{s}_j} \mathcal{E}_\mathbf{i} $. Here, $\mathbf{i} $ is a qubit in the slice $\mathbf{s}_j$ and
\begin{equation}
 \mathcal{E}_\mathbf{i} (\hat{\rho}) = (1- \sum_{\hat{O}=\hat{X},\hat{Y},\hat{Z}}p_{\hat{O}})\hat{\rho} + \sum_{\hat{O}=\hat{X},\hat{Y},\hat{Z}} p_{\hat{O}} \hat{O_\mathbf{i} }\hat{\rho}\hat{O}_\mathbf{i} ^\dagger,
\label{eq:Error}
\end{equation}
with $\hat{X},\hat{Y},\hat{Z}$ being the standard Pauli matrices and $0 \leq p_{\hat{O}} \leq 1$ the probability of a single-qubit error $\hat{O}$, with $\sum_{\hat{O}=\hat{X},\hat{Y},\hat{Z}}p_{\hat{O}} \leq 1$. The channel describing the noisy preparation of the state $\hat{\rho}_\text{out}$ from the initial state $\hat{\rho}_\text{in} = \ket {\Psi_\text{in}}\bra{\Psi_\text{in}}$ is obtained by alternating the unitary slice operator $\mathcal{U}_{j, j+1} (\hat{\rho}) = \hat{U}_{\mathbf{s}_{j},\mathbf{s}_{j+1}} \hat{\rho} \hat{U}^\dagger_{\mathbf{s}_{j}, \mathbf{s}_{j+1}}$ and the error channel $\mathcal{E}_j (\hat{\rho})$,
\begin{equation}
    \hat{\rho}_\mathrm{out} = \mathcal{U}_{L-1, L} \mathcal{E}_{{L-1}} \cdots \mathcal{E}_{3}\mathcal{U}_{2, 3}\mathcal{E}_{2} \mathcal{U}_{1, 2}(\hat{\rho}_\mathrm{in}).
    \label{eq:NoisePrep}
\end{equation}
While $\hat{\rho}_{\text{out}}$ will invariably differ from $\ket{\Psi_{\text{out}}}\bra{\Psi_{\text{out}}}$, our interest is in whether $\hat{\rho}_{\text{out}}$ can exhibit the same class of ordering as the noiseless target state---that is, do these states belong to the same \textit{mixed-state phase} \cite{sang_mixed-state_2026, ellison_towards_2025, fan_diagnostics_2024}? See App.~A for a precise discussion of this notion.

A few comments are in order: First, the noise process considered here has to be contrasted with local incoherent noise on a pure state $\ket{\Psi_\text{out}}$, which symmetry-breaking or topological order is stable against below a threshold error probability $p_c$~\cite{dennis2002topological, wang2003confinement, bombin_strong_2012, katzgraber_error2009, breuckmann2016localdecoders, kubica_three2018, chubb2021statistical, fan_diagnostics_2024}: In Eq.~\eqref{eq:NoisePrep}, gates $\hat{U}_{\mathbf{s}_{j},\mathbf{s}_{j+1}}$ can propagate an originally localized error, potentially leading to a non-local perturbation which would destroy any long-range order or entanglement of the state $\ket{\Psi_\text{out}}$ even at infinitesimal error probability $p$. This prompts the need for sensible gate design protecting against uncontrolled error spread. Errors on the second slice $\mathbf{s}_{j+1}$ in each step map to local errors and thus need not be considered, see SM~\cite{SM} Sec.~B. Second, the stability of a sequential generation is distinct from stability of a given order to thermal fluctuations~\cite{bravyi2009no, castelnovo_topological_2008, alicki_thermalstabilitiy2010} as shown in Tab.~\ref{tab:stability}.

\textbf{\emph{Preparation of 3D symmetry-breaking order.---}} As an instructive example, we will first consider the sequential generation of a state exhibiting symmetry breaking in 3D as illustrated in Fig.~\ref{fig:fig1}. We consider a cubic lattice with a single qubit on every vertex, and choose the mixed target state
\begin{equation}
    \hat{\rho}_{\text{SSB}} = \frac{1}{2}\left( \vert 0 \rangle\langle 0 \vert^{\otimes N} + \vert 1 \rangle \langle 1 \vert^{\otimes N}\right).
    \label{eq:rhoSSB}
\end{equation}
The symmetry-breaking long-range order of this state can be detected via the connected correlator $C_{\mathbf{i} , \mathbf{j} } = \langle \hat{Z}_{\mathbf{i} } \hat{Z}_{\mathbf{j} }\rangle - \langle \hat{Z}_{\mathbf{i} }\rangle \langle \hat{Z}_{\mathbf{j} }\rangle$ being finite as $|\mathbf{i} - \mathbf{j}|\to \infty$. We refer to this state as having spontaneous symmetry breaking (SSB).

In the absence of any Pauli error, this state can be prepared using a generalization of the preparation procedure described in the introduction for the 1D GHZ state: Choosing the sequential circuit to act along the fully diagonal (1,1,1) axis of the cubic lattice (relative to the lattice vectors $\mathbf{e}_i$, see lattice and coordinate system in Fig.~\ref{fig:fig1}a), each layer $\hat{U}_{\mathbf{s}_{j}, \mathbf{s}_{j+1}}$ can be decomposed into CNOT gates, the target qubits of which are vertices in the layer $\mathbf{s}_{j+1}$, with any one of its three neighbors in the preceding layer $\mathbf{s}_{j}$ as the control qubit. Starting from a boundary which is a probabilistic mixture, $\hat{\rho}_b = \frac{1}{2}\left( \vert 0 \rangle\langle 0 \vert_{\mathbf{s}_1} + \vert 1 \rangle \langle 1 \vert_{\mathbf{s}_1}\right)$, this circuit copies the classical bit encoded in the boundary state into the entire system, layer-by-layer, thereby preparing the state Eq.~\eqref{eq:rhoSSB}. Indeed, in this framework the same circuit could also prepare the long-range entangled GHZ state $\hat{\rho}_{\text{GHZ}} = \ket{\text{GHZ}}\bra{\text{GHZ}}$, with $\vert \text{GHZ}\rangle = \frac{1}{\sqrt{2}}\left( \vert 0 \rangle^{\otimes N} + \vert 1 \rangle^{\otimes N}\right)$, provided the boundary state can already be put into this superposition, $\ket{\Psi_b} = \ket{\text{GHZ}}_{\mathbf{s}_1}$.

How do the Pauli errors in Eq.~(\ref{eq:NoisePrep}) affect the output state for this preparation protocol?
Since $\hat{Z}$ errors for $p_{\hat{Z}}\neq 0$ commute with the gates $\hat{U}_{\mathbf{s}_{j},\mathbf{s}_{j+1}}$, they remain local under the circuit and therefore act only as local noise on the prepared state. While this destroys the phase coherence and thus the long-range entanglement of the state $\hat{\rho}_{\text{GHZ}}$, the long-range order of the state $\hat{\rho}_{\text{SSB}}$ is left untouched. For the purposes of preparing the symmetry-broken state, such errors therefore need not be treated. By contrast, $\hat{X}$ and $\hat{Y}$ errors induce bit flips, which are propagated indefinitely through the system by the circuit. They are thus mapped to non-local bit-flip noise on the physical state, which destabilizes the SSB order.

This problem can be alleviated by using the geometry of the underlying lattice. Instead of preparing a qubit $\mathbf{i}$ conditioned on a single neighbor in the preceding slice, we can perform a majority vote on all three of its nearest neighbors $\mathbf{j} ,\mathbf{k} ,\mathbf{l} $ in this slice, as implemented by the unitary
\begin{equation}
\begin{split}
    \hat{U}_{\text{SSB}}\vert \sigma_\mathbf{j} \sigma_\mathbf{k} \sigma_\mathbf{l} \rangle_{\mathbf{s}_j} \vert \mu_{\mathbf{i}}\rangle_{\mathbf{s}_{j+1}}=\vert \sigma_\mathbf{j} \sigma_\mathbf{k} \sigma_\mathbf{l} \rangle_{\mathbf{s}_j} \vert \mu_{\mathbf{i}} \oplus \sigma_{\text{MAJ}} \rangle_{\mathbf{s}_{j+1}} ,\\ \sigma_{\text{MAJ}}  \coloneqq  \begin{cases}
        0, \, \text{if}\, \sigma_\mathbf{j} + \sigma_\mathbf{k} + \sigma_\mathbf{l}  \le 1 \\
        1, \, \text{otherwise}
    \end{cases}.
\end{split}
\label{eq:MajorityRule}
\end{equation}
With the slice $\mathbf{s}_{j+1}$ initialized in the state $\ket{0}_{\mathbf{s}_{j+1}}$, this gate copies the majority spin $\sigma_{\text{MAJ}}$ into qubit $\mathbf{i}$. Concretely, this unitary can be constructed from three Toffoli gates acting on the qubit with all possible pairs of its neighbors as control qubits. The slice unitary $\hat{U}_{\mathbf{s}_{j}, \mathbf{s}_{j+1}}$ is built from three layers of $\hat{U}_{\text{SSB}}$ gates, in the geometry shown in Fig.~\ref{fig:fig1}b. 

To understand how errors propagate through this modified gate, observe that the syndromes of the stabilizers of $\hat{\rho}_{\text{SSB}}$ are domain walls separating regions of 0s and 1s; in three dimensions, these form closed surfaces. At a given time $t$, the marginal state of the preceding layer $\mathbf{s}_{t}$ is a two-dimensional slice through this configuration, whose syndrome will consist of closed loops. Crucially, these loops have a curvature which we can use to locally identify which side of the domains  is in the erroneous spin state; this is achieved by the majority rule. The unitary gate $\hat{U}_{\text{SSB}}$ iteratively shrinks loops according to their curvature, preventing the uncontrolled spread of the wrong domain. This curvature-driven correction mechanism is illustrated schematically in Fig.~\ref{fig:fig1}c.

As a result, we expect that the typical syndromes of the noise-corrupted state $\hat{\rho}_{\text{out}}$ [Eq.~\eqref{eq:NoisePrep}] will be made up of small closed domain walls, for bit-flip rates $p_{\hat{X}}+p_{\hat{Y}}$ below a finite threshold. The small ``wrong" domains can be removed by a local correction channel to $\hat{\rho}_{\text{out}}$, and so the noise-corrupted state is local channel-connected to the target $\hat{\rho}_{\text{SSB}}$. In this sense, our circuit generates SSB order in a noise-robust way.

To test this expectation quantitatively, we simulate the stable generation circuit using a classical cellular automaton that implements the majority rule in a system of size $L \times L \times 2L$, where the final direction is the time-like $z$ direction of the sequential circuit (the $(1,1,1)$ direction relative to the lattice vectors $\mathbf e_i$) and the first two directions $x$ and $y$ are orthogonal to it with periodic boundary conditions. We initialize the first slice in the all-$\ket{0}$ or all-$\ket{1}$ state with probability 1/2 each, and  stochastically add bit-flips to the control layers of the sequential preparation with a probability $p_{\hat{X}}$. Since $\hat{Y}=i\hat{X}\hat{Z}$, $\hat{Y}$ errors contribute the same bit-flip component and are therefore captured by the same stochastic process. To probe the bulk order of the resulting state, we compute the connected two-point correlations $C_{\mathbf{i} , \mathbf{j} } = \langle \hat{Z}_{\mathbf{i} } \hat{Z}_{\mathbf{j} }\rangle - \langle \hat{Z}_{\mathbf{i} }\rangle \langle \hat{Z}_{\mathbf{j} }\rangle$.

\begin{figure}
    \centering
    \includegraphics[width=1\linewidth]{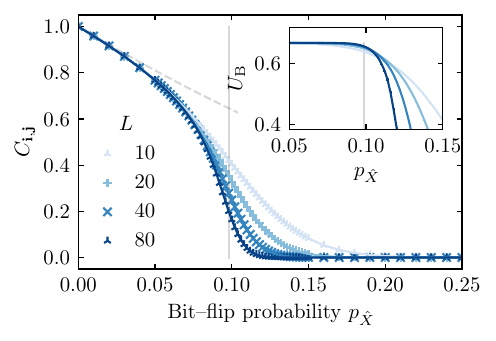}
    \caption{Connected two-site correlation function $C_{\mathbf{i} , \mathbf{j} } = \langle \hat{Z}_{\mathbf{i} } \hat{Z}_{\mathbf{j} }\rangle - \langle \hat{Z}_{\mathbf{i} }\rangle \langle \hat{Z}_{\mathbf{j} }\rangle$ for a sequentially prepared state using the local unitary $\hat{U}_{\text{SSB}}$, Eq.~(\ref{eq:MajorityRule}), for different linear system sizes $L$ and bit-flip probabilities $p_{\hat{X}}$. The sites $\mathbf{i}$ and $\mathbf{j}$ are in the bulk of the system and separated by $L$ in the $z$ direction, corresponding to the direction of the sequential circuit, $\mathbf{j} = \mathbf{i}  + L \cdot \hat{e}_z$. For error rates below the extracted critical probability of $p^\mathrm{c}_{\hat{X}} \approx 0.1$ (gray vertical line), the correlations stay finite in the thermodynamic limit, indicating the survival of long-range order. At very low error rates, the correlations follow the prediction of independent local noise (dashed line). Inset: Binder cumulant $U_B$ of the magnetization, serving as an order parameter. We see a second-order phase transition from a state with long-range correlations to a trivial paramagnet.}
    \label{fig:bulk_correlation.pdf}
\end{figure}

We plot $C_{\mathbf{i} , \mathbf{j} }$ in Fig.~\ref{fig:bulk_correlation.pdf}, for $\mathbf{i} $ and $\mathbf{j} $ separated by half the system size in the $z$ direction. At $p_{\hat{X}} = 0$, the state $\hat{\rho}_{\text{SSB}}$ is generated with $C_{\mathbf{i} , \mathbf{j} } = 1$. These long-range correlations persist up to a threshold of $p^c_{\hat{X}} \approx 0.1$ (estimated using the Binder cumulant $U_\mathrm{B} = 1-{\langle  \hat{M} ^4 \rangle}/{3\langle  \hat{M}^2\rangle^2}$~\cite{binder_finite_1981, binder_critical_1981}, where $\hat{M} = \frac{1}{N}\sum_{i = 1}^N \hat{Z}_i$ is the magnetization). At low error probabilities, the correlation function is well-approximated by single-qubit noise on the physical level, $C(p_{\hat{X}}) =  1-4 p_{\hat{X}} + \mathcal{O} (p_{\hat{X}}^2)$; deviations appear closer to the threshold as the bit-flips can spread slightly before being shrunken to zero by the majority rule. 
This indicates the circuit can indeed successfully correct the sparsely appearing errors and preserve the long-range order up to $p^c_{\hat{X}}$.
\begin{figure}
    \centering
    \includegraphics[width=0.95\linewidth]{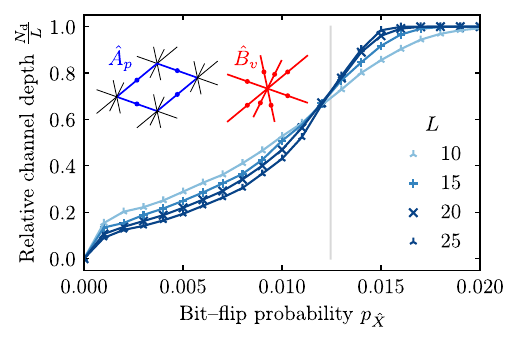}
    \caption{Mean relative depth ${N_d}/{L}$ of the local channel connecting the sequentially prepared state $\hat{\rho}_{\text{out}}$ using the local unitary $\hat{U}_{(3,1)\text{TC}}$, Eq.~\eqref{eq:TCUnitary}, to the error-free (3,1) toric code wavefunction $\ket{(3,1)\text{TC}}$, for different linear system sizes $L$ and bit-flip probabilities $p_{\hat{X}}$. Below the extracted critical probability of $p^\mathrm{c}_{\hat X} \approx 0.013$ (gray vertical line), the depth of the channel scales sublinearly with $L$, indicating long-range entanglement and topological order in the noisy state $\hat{\rho}_{\text{out}}$. Inset: Visualization of the stabilizers. The stabilizer $\hat{A}_p$ consists of $\hat{Z}$ operators on the four edges surrounding a plaquette $p$ (blue). The stabilizer $\hat{B}_v$ consists of $\hat{X}$ operators on the eight edges adjacent to a vertex $v$ (red).}
    \label{fig:relative_unitary_circuit_depth.pdf}
\end{figure}

\textbf{\emph{4D Toric Code.---}} As discussed in the preceding section, any phase-flip error leads to loss of coherence in the GHZ state, thus precluding the generation of a long-range entangled state through a noisy circuit. However, we can also apply the logic presented above to topologically ordered stabilizer states, which feature robust long-range entanglement~\cite{kitaev_faulttolerant2003, kitaev_topological_2006}.

Concretely, we consider the four-dimensional $(3,1)$ toric code state, which is a higher-dimensional generalization of the more familiar 2D toric code; it is defined on a 4D hypercubic lattice with qubits located on the edges. Its stabilizer groups are generated by products of $\hat{Z}$ operators on the qubits on the edges of a plaquette $p$, $\{\hat{A}_p = \prod_{\mathbf{i} \in \partial p} \hat{Z}_\mathbf{i}  \vert \forall p\}$, and products of $\hat{X}$ operators acting on the qubits on the edges adjacent to a vertex $v$, $\{\hat{B}_v = \prod_{\mathbf{i} \in \delta v} \hat{X}_\mathbf{i}  \vert \forall v\}$, where $\partial p$ is the boundary of $p$ and $\delta v$ the co-boundary of $v$; see inset of Fig.~\ref{fig:relative_unitary_circuit_depth.pdf} for a visualization of the stabilizers. The ground state $\ket{(3,1)\text{TC}}$ of the associated stabilizer Hamiltonian $\hat{H}_{(3,1)\text{TC}} =-\sum_{p} \hat{A}_p - \sum_{v} \hat{B}_v $ is an equal-weighted superposition of all closed loops in $\hat{X}$ basis, or alternatively all closed volumes in $\hat{Z}$ basis. The stabilizer groups form a $3$-form and a $1$-form symmetry, respectively~\cite{gaiotto_generalized_2015, mcgreevy2023generalized}. The syndromes corresponding to violations of the $3$-form symmetry induced by the $Z$ stabilizers form closed membranes, while those associated with violations of the $1$-form $X$-type symmetry are point-like. Importantly, this state and Hamiltonian should be contrasted with the more commonly discussed self-dual 4D $(2,2)$ toric code with two 2-form symmetries, whose topological order exhibits thermal stability~\cite{dennis2002topological, alicki_thermalstabilitiy2010} (c.f. Tab.~\ref{tab:stability}).

The closed-membrane structure of the 3-form syndromes is reminiscent of the 3D domain walls in the symmetry-breaking example and will similarly allow the unitaries in the sequential generation to locally detect and counteract the corresponding errors. We consider a hyper-cubic lattice of size $(L_x, L_y, L_z, L_w)$, where the $w$ direction points along the fully diagonal (1,1,1,1) axis relative to the lattice vectors $\mathbf{e}_i$ and serves as the direction the preparation circuit acts on sequentially. The other three directions $x,y,z$ are orthogonal to $w$ and have periodic boundary conditions. Each three-dimensional hyper-layer $\mathbf{s}_j$ of the sequential circuit is orthogonal to the $w$ direction, with the first layer initialized in a product state, $\ket{\Psi_b} = \ket{0}^{\otimes L_xL_yL_z}_{\mathbf{s}_1}$.

The layer unitary $\hat{U}_{\mathbf{s}_{j}, \mathbf{s}_{j+1}}$ of the sequential circuit is built from gates $\hat{U}_{(3,1)\text{TC}}$, each acting on 16 neighboring qubits. The gate has a controlled-unitary structure, with twelve qubits $\sigma_1,\dots ,\sigma_{12}$ in layer $\mathbf{s}_j$ as controls, and four qubits $\mu_1,\dots,\mu_4$ in layer $\mathbf{s}_{j+1}$ as targets. Between each other, these 16 qubits share six plaquettes (for a more detailed description of the support and action of the unitary, see supplemental material~\cite{SM} Sec.~C). For each state $\ket{\sigma_1\cdots \sigma_{12}}$, the unitary is chosen such that, when it acts on the state $\ket{\mu_1\cdots \mu_4}= \ket{0000}$, the resulting output violates the fewest stabilizers. Specifically,

\begin{equation}
    \hat{U}_{(3,1)\text{TC}} \ket{\sigma_1\cdots \sigma_{12}}\ket{0000} =\frac{1}{\sqrt{2}} \ket{\sigma_1\cdots \sigma_{12}} \sum_{\substack{\{\mu_1,\dots\mu_4\} \\ \in C_{\sigma_1\cdots \sigma_{12}}}}  \ket{\mu_1\cdots\mu_4},
    \label{eq:TCUnitary}
\end{equation}
where for each control configuration $\sigma_1, \ldots, \sigma_{12}$, the set $C_{\sigma_1\cdots \sigma_{12}}$ contains two four-qubit states that are related by a complete spin flip, namely $\ket{\mu_1\cdots\mu_4}$ and $\ket{\mu^\prime_1\cdots\mu^\prime_4} = \left( \prod_{i=1}^4 \hat X_{i} \right) \ket{\mu_1\cdots\mu_4}$. The sets $C_{\sigma_1\cdots \sigma_{12}}$ are chosen such that the minimum number of plaquette stabilizers $\hat {A} _p$ across the sixteen qubits is violated, conditioned on the values of $\sigma_1, \ldots, \sigma_{12}$. This is analogous to the majority-vote choice for the SSB state, Eq.~\eqref{eq:MajorityRule}. By taking a uniform superposition of $\ket{\mu_1\cdots\mu_4}$ and $\ket{\mu^\prime_1\cdots\mu^\prime_4}$,  we also ensure that the vertex stabilizer $\hat{B}_v$ is satisfied. We note that there may be multiple choices of pairs of states with the same number of $\hat{A}_p$ violated---in this case, a particular choice for $C_{\sigma_1, \ldots, \sigma_{12}}$ is made in advance, see SM~\cite{SM} Sec.~C. In the numerics described later, we select this prescription randomly for each instance of the gate $\hat U_{(3,1)\text{TC}}$; alternatively, a layered construction can be used where the layers $\mathbf s_j$ alternate periodically between the three different prescriptions.

This circuit is designed to prevent the proliferation of errors when noise is added. As in the 3D SSB case, a $\hat{Z}$-type error arising during a given slice $\mathbf{s}_{j}$ commutes through the gate $\hat{U}_{(3,1)\text{TC}}$, and so these processes are equivalent to the application of noise after the whole state is prepared. The topological order of the toric code is known to be stable to this type of noise, up to some non-zero critical threshold of $p_{\hat{Z}}$. Indeed, typical vertex ($\hat{B}_v$) syndromes of $\hat{\rho}_{\text{out}}$ will feature tightly-bound pairs of point-like excitations, which can be removed by a local channel after state preparation.

Thanks to our majority-vote-like construction, the topological order is also robust against $\hat{X}$ and $\hat{Y}$ errors that arise during the sequential circuit. Within a given three-dimensional time slice, violations of the $\hat{A}_p$ stabilizers form loops, which are systematically shrunk in successive layers by the gates $\hat{U}_{(3,1)\text{TC}}$. Below some finite threshold of $p_{\hat{X}}+p_{\hat{Y}}$, the noise-corrupted state $\hat{\rho}_{\text{out}}$ will feature small membrane-like clusters of plaquette ($\hat{A}_p$) syndromes, which can also be removed by a local channel.

To illustrate this, we simulate the action of our unitary circuit in the presence of noise using a classical cellular automaton. In this case, the classical simulation is possible because the observable is a function of the syndrome associated with one set of stabilizers, that is, the plaquette ($\hat{A}_p$) stabilizers, and is linear in the ensemble. We consider a system of size $L\times L \times L \times 2L$, with the automaton implementing the rule of Eq.~(\ref{eq:TCUnitary}) and interspersed random bit-flip noise, corresponding to $p_{\hat{X}} > 0$. From each generated classical configuration, we can infer a $Z$-type syndrome for the physical state $\hat{\rho}_{\text{out}}$. On this syndrome, we then run a local cellular automaton decoder, which shrinks corners of membranes in $w$ direction via a local rule~\cite{kubica_cellular-automaton_2019} (see also SM~\cite{SM} Sec.~D). If this decoder succeeds in removing syndromes within a subextensive time, then there exists a channel of correspondingly low depth that can bring $\hat{\rho}_{\text{out}}$ into the subspace stabilized by all $\hat{A}_p$. Combined with our prediction that $\hat{B}_v$ violations can be removed locally, this implies local channel connectivity from $\hat{\rho}_{\text{out}}$ to $\hat{\rho}_{\text{(3,1)TC}}$. This in turn implies that $\hat{\rho}_{\text{out}}$ must exhibit nontrivial topological order; see App.~A.

In Fig.~\ref{fig:relative_unitary_circuit_depth.pdf}, we plot the average depth $N_{\text{d}}$ at which this decoder removes the syndrome, normalized by linear system size $L$. Below a noise threshold of $p^c_{\hat{X}} \approx 0.013$, this depth scales sublinear in $L$; this indicates that the generated mixed state $\hat{\rho}_{\text{out}}$ can be connected to the topologically ordered fixed point $\hat{\rho}_{(3,1)\text{TC}} = \ket{(3,1)\text{TC}}\bra{(3,1)\text{TC}}$ by a channel of subextensive depth. In this regime, we also expect intuitively that the converse relation holds, i.e.,~that the noisy state $\hat{\rho}_{\text{out}}$ can  be reached from the fixed point $\hat{\rho}_{(3,1)\text{TC}}$ via a quasi-local channel. This is because typical syndromes of $\hat{\rho}_{\text{out}}$ feature small localized clusters (See SM~\cite{SM} Sec.~E for numerical data on the maximal cluster size). This two-way connectivity implies that $\hat{\rho}_{\text{out}}$ and $\hat{\rho}_{\text{(3,1)TC}}$ are in the same phase; see App.~A.

\textbf{\emph{General mechanism for stable sequential state generation. ---}} More generally, states associated with topological Calderbank-Shor-Steane (CSS) codes can be generated via a unitary sequential circuit in a way that is stable against weak Pauli noise, provided one class of syndromes (corresponding to the $\hat Z$ stabilizers) forms closed manifolds of dimension $d = 2$ (closed membranes), or higher. 
These states are characterized by a $p$-form symmetry with $p \geq 3$, including symmetry-breaking states in dimensions $D \geq 3$ \footnote{Here, we interpret symmetry breaking in $D$ dimensions as an anomaly between a $D$-form symmetry and a weak $0$-form symmetry~\cite{liu_detecting_2025}.} and topologically ordered states with $D \geq 4$. The unitary gates in these circuits prepare a layer of qubits by using the preceding layer as control input, ensuring that all stabilizers are satisfied. Errors of type $\hat{Z}$ in the build-up are accounted for by design, as they commute with the unitaries and map to local noise, which does not violate local stabilizers (for SSB order) or does not destroy topological order below a threshold error rate. By contrast, bit-flip errors generated by $\hat{X}/\hat{Y}$ operators on the control qubits can proliferate and destroy the order in principle. The local gates can use the curvature of a closed $(d-1)$ dimensional syndrome configuration in the current time slice to shrink it iteratively via a majority vote of the implicated stabilizers, as shown in Fig.~\ref{fig:fig1}. For low error rates, these syndromes should thus only appear as closed $d$ dimensional objects of finite size, which can be cleaned up by a local correction channel.
Despite the stabilizer nature of the state, this local shrinking process requires that the unitary gate cannot be decomposed into Clifford gates, which lose information in the presence of errors exponentially fast~\cite{nelson_nonclifford_2025}. Intuitively, the majority-vote rule necessitates multi-control gates such as Toffoli gates, which cannot be built exclusively from Clifford gates; in App.~B, we provide a precise argument why enforcing the Clifford property on our local gate leads to a contradiction.

The shrinking process can also be viewed as holographic error correction in each time step; using the local curvature, this can be implemented directly in the unitary gates. If one allows for stabilizer measurements and feedback within each slice, states with $d=1$ dimensional syndrome patterns (closed loops) can be prepared sequentially, including 2D symmetry-breaking order or topological states with a $2$-form symmetry in dimensions $D \geq 3$, such as the 3D toric code~\cite{castelnovo_topological_2008, williamson_stability_2021, delcamp_tensor_2021}, or the 4D (2,2) loop toric code~\cite{dennis2002topological, breuckmann2016localdecoders} (see also Tab.~\ref{tab:stability}). The stabilizer violations that can be detected within a given slice are isolated, pairwise created points whose position can be obtained via stabilizer measurements. A decoder such as Minimum Weight Perfect Matching can predict how to pair them up within the slice~\cite{Barahona1982Morphology, galil_efficient1986}; a local unitary is then chosen which moves each error closer to its partner. For low error probabilities, this prevents proliferation of the errors.

\textbf{\emph{Conclusion. ---}} We have proposed a general framework to construct sequential quantum circuits for stabilizer states with non-trivial order which remain stable to Pauli noise, based on the local curvature of the syndrome patterns. As symmetry-broken states and topologically ordered states can be seen as instances of classical and quantum memory respectively, these protocols promise a novel pathway towards systems which can store quantum information.

While we have focused on Pauli noise in this work, the applicability of the circuits is, in fact, more general. Biased noise, against which isotropic majority-rule circuits for long-range order fail, can be treated by circuits adapting more sophisticated classical automata such as Toom's rule~\cite{toomsrule}. In Sec.~F of the supplement~\cite{SM}, we provide such an improved unitary circuit explicitly. On the other hand, restricting to strongly symmetric noise allows for preparation of long-range entangled states which are strongly symmetric such as $\hat{\rho}_{\text{GHZ}}$, in contrast to weakly symmetric states such as $\hat{\rho}_{\text{SSB}}$ which can be generated without this restriction~\cite{victor_symmetries2014, buvca2012note, ziereis_strong--weak_2025}. Furthermore, for topologically ordered states one can consider general non-Pauli noise; we would expect the circuit to still be stable, as such an error can be decomposed into Pauli strings, similar to error discretization in standard quantum error correction~\cite{raussendorf_key_2012}.

The existence of a noise-resistant circuit for a fixed point is intricately linked to its tensor network representation and its virtual subspaces~\cite{shukla_boson_2018, williamson_stability_2021}; for a more detailed discussion, see SM~\cite{SM} Sec.~A. While we have focused on stabilizer states in this work, a natural generalization would consider more general topological orders in various dimensions, such as string-net models and their higher-dimensional analogs~\cite{levin_stringnet2005,soejima_isometric_2020, zhou_finite_2025}, or fracton order~\cite{haah_local_2011, vijay_fracton_2016, prem_cage-net_2019}, which also have well-understood fixed points. Furthermore, the tensor network perspective allows for finite-correlation deformations of these fixed points whose topological order persists, while still directly providing the local unitary gates of the sequential generation circuit~\cite{liu_simulating2024, boesl_quantum2025, boesl2025skeleton}.

\textit{\textbf{Acknowledgments.---}} We acknowledge support from the Deutsche Forschungsgemeinschaft (DFG, German Research Foundation) under Germany’s Excellence Strategy--EXC--2111--390814868, TRR 360 – 492547816 and DFG grants No. KN1254/1-2, KN1254/2-1, the European Union (grant agreement No 101169765), as well as the Munich Quantum Valley, which is supported by the Bavarian state government with funds from the Hightech Agenda Bayern Plus. M.M.~acknowledges support from Trinity College, Cambridge.

\textit{\textbf{Data availability.---}}Numerical codes are available upon reasonable request on Zenodo~\cite{zenodo}. 

\section*{End Matter}

\section{Appendix A: Definition of stability of sequential circuit and relationship with mixed-state topological order\label{app:mixed}}
In the main text, we described the sequential circuits for the 3D symmetry-broken state and the 4D (3,1) toric code state as being stable to noise in a particular sense, which we elaborate on here.

Similar to the case where noise is applied to an already prepared noiseless state (which is the case most often considered in the literature), the output of the noisy sequential circuit $\hat{\rho}_{\rm out}$ [Eq.~\eqref{eq:NoisePrep}] will inevitably differ somewhat from the target state. Accordingly, our aim in constructing a stable sequential circuit should not be to achieve $\hat{\rho}_{\text{out}} \approx \hat{\rho}_{\rm target}$. Instead, we wish to ensure that $\hat{\rho}_{\rm out}$ exhibits the same kind of order as the target state, in the sense of being in the same \textit{mixed-state phase}.

For the purposes of this work, we will formally consider two states $\hat{\rho}_1$, $\hat{\rho}_2$ to be in the same mixed-state phase if there exist channels $\mathcal{N}_{2 \leftarrow 1}$, $\mathcal{N}_{1 \leftarrow 2}$ such that the following two conditions are met.
\begin{enumerate}
    \item $\mathsf{D}(\hat{\rho}_2, \mathcal{N}_{2 \leftarrow 1}[\hat{\rho}_1]) \leq \epsilon(L)$ and $\mathsf{D}(\hat{\rho}_1, \mathcal{N}_{1 \leftarrow 2}[\hat{\rho}_2]) \leq \epsilon(L)$ for some small $\epsilon(L) = o(1)$, where $\mathsf{D}(\hat{\rho}, \hat{\sigma})$ is some reasonable measure of distance between mixed states $\hat{\rho},\hat{\sigma}$ (e.g.~trace distance).
    \item $\mathcal{N}_{2 \leftarrow 1}$ and $\mathcal{N}_{1 \leftarrow 2}$ can each be implemented by a circuit of depth $d \leq f(L)$ for some function $f(L) \leq o(L)$, assisted by local ancilla qubits that are traced out at the end. Each gate in these circuits acts only on system and ancilla qubits within some constant range.
\end{enumerate}
Here, we denote with $o(L)$ a strictly smaller order of magnitude than $L$, i.e., $\lim_{L\to \infty }o(L)/L = 0$. Note that the ancilla-assisted circuits in 2. include local channel circuits \cite{mcginley2025lower}. Thus in the main text, we refer to connectivity of this type as being via a `low-depth channel circuit'. Roughly speaking, since channels of this form can only remove (but cannot create) long-range correlations and topological order, the existence of such a map $\mathcal{N}_{2 \leftarrow 1}$ implies that $\hat{\rho}_1$ has `at least as much order' as $\hat{\rho}_2$. We note that more refined notions of mixed-state phase equivalence do exist, where the channels must also be locally reversible~\cite{sang2025mixedstatephaseslocalreversibility}; these are needed to capture e.g.~classical states that exhibit strong-to-weak symmetry breaking. However, the above definition is simple, operationally well-motivated, and suffices to distinguish topological, symmetry-breaking, and trivial orders.

Our numerical experiments on the 4D (3,1) toric code indicate that the noise-corrupted state $\hat{\rho}_{\rm out}$ can be converted to the clean target state $\hat{\rho}_{(3,1)\text{TC}}$ by means of stabilizer measurements and a sub-extensive number of rounds of \textit{local} feedback. Each measurement and local feedback step can be written as a single layer of a local channel circuit. Accordingly, our claim implies the existence of a low-depth channel circuit that maps $\hat{\rho}_{\rm out}$ to $\hat{\rho}_{(3,1)\text{TC}}$. 

Technically, we have not demonstrated connectivity in the reverse direction, i.e.~from $\hat{\rho}_{(3,1)\text{TC}}$ to $\hat{\rho}_{\rm out}$. In the usual case where one considers local noise acting on a clean state at the physical level, this direction would be immediate, since the noise channel itself would constitute the low-depth circuit connecting $\hat{\rho}_{(3,1)\text{TC}}$ to $\hat{\rho}_{\rm out}$. However, it is not necessarily true that the effect of the mid-circuit noise in Eq.~\eqref{eq:NoisePrep} is equivalent to the action of a local channel on the physical level. Nevertheless, we intuitively expect that for noise rates below threshold, where the syndromes generated form in small local clusters, the distribution of these clusters does not exhibit long-range correlations, and could be mimicked by some effective low-depth local channel circuit.

Regardless, the existence of a low-depth local channel circuit connecting the noise-corrupted state $\hat{\rho}_{\rm out}$ to the clean state $\hat{\rho}_{(3,1)\text{TC}}$ establishes the former as being at least as ordered as $\hat{\rho}_{(3,1)\text{TC}}$, in the sense defined above. Concretely, it follows that $\hat{\rho}_{\rm out}$ is a non-trivial state, in that it cannot be generated by a low-depth channel circuit starting from a product state. We therefore use this as an operational signature of mixed-state topological order in this work.

\section{Appendix B: Non-Clifford Condition on Error-Correcting Gates\label{app:clifford}}
We will show that the concrete examples in the main text for sequential preparation circuits that are stable against probabilistic Pauli noise cannot be decomposed exclusively into Clifford gates, even though the target states are stabilizer states and can be prepared via Clifford circuits in the absence of noise. In this section, we provide a direct argument why the tensor-network inspired circuits (see SM~\cite{SM} Sec. A) for CSS codes cannot be stable against Pauli noise and be Clifford at the same time provided non-trivial order is to be generated.

Our preparation protocol sequentially entangles two adjacent slices $\mathbf{s}_j$ and $\mathbf{s}_{j+1}$ in each time step. The unitary acting on these two slices can be further decomposed into local unitary gates. For simplicity, we consider the same gate $\hat{U}$ at different positions in the two slices, as is the case for the examples in the main text. The first slice $\mathbf{s}_j$ is supposed to be already prepared and serves as a control input in the $\hat{Z}$ basis, while the second slice $\mathbf{s}_{j+1}$ is assumed to be in the product state $\ket{0\cdots 0}$.
The relevant action of the unitary $\hat{U}$ is
\begin{equation}
    \hat{U}\ket{\mathbf{x}}_{\mathbf{s}_{j}}\ket{0 \cdots 0}_{\mathbf{s}_{j+1}} = \ket{\mathbf{x}}_{\mathbf{s}_{j}} \ket{\psi_{\mathbf{x}}}_{\mathbf{s}_{j+1}},
    \label{eq:UGeneral}
\end{equation}
where $\ket{\mathbf{x}}_{\mathbf{s}_{j}}$, with $\mathbf{x} \in \{0,1 \}^n$, is a $\hat{Z}$ basis state of the $n$ qubits in the support of $\hat{U}$ in slice $\mathbf{s}_j$ and $\ket{\psi_{\mathbf{x}}}_{\mathbf{s}_{j+1}}$ a superposition of certain $Z$ basis states in the support of $U$ in slice $\mathbf{s}_{j+1}$. In particular, there is a subset $G$ of ``valid" input configurations $\ket{\mathbf{x}}_{\mathbf{s}_{j}}$ such that $\ket{\mathbf{x}}_{\mathbf{s}_{j}} \ket{\psi_{\mathbf{x}}}_{\mathbf{s}_{j+1}}$ respects all $\hat{Z}$ stabilizers which lie in the support of $\hat{U}$; the amplitudes of the associated $\ket{\psi_{\mathbf{x}}}_{\mathbf{s}_{j+1}}$ can be chosen such that all $\hat{X}$ stabilizers are satisfied as well.

We only consider noise on the first slice (see SM~\cite{SM} Sec. B). We want the Pauli error appearing before the application of the gate to map to a local physical error. For $\hat{Z}$ noise, this is the case by construction as any qubit $\mathbf{i} \in \mathbf{s}_j \cap \text{supp}(\hat{U})$ only serves as control, so $[\hat{U}, \hat{Z}_{\mathbf{i}}] = 0$. In general, $\hat{X}$ and $\hat{Y}$ errors propagate by such a gate. We thus impose this commutation property on the subspace $G$ of valid configurations as a condition for stability:
\begin{equation}
    [\hat{U}, \hat{X}_{\mathbf{i}}]\hat{P}_G = [\hat{U}, \hat{Y}_{\mathbf{i}}]\hat{P}_G = 0,
    \label{eq:FlipCommutation}
\end{equation}
where $\hat{P}_G$ is the projector on the valid input configurations $G$. This condition implies that the Pauli error commutes in the subset G with the preparation unitary $\hat U$ and therefore stays local. The motivation behind this definition is that it implies that a majority vote on the stabilizers in the support of $\hat U$ gives a definite answer if the control configuration is outside of $G$ by only one bit-flip; if the local closed loop of violated stabilizers in the slice $s_{\mathbf j}$ is strictly convex, this procedure shrinks it iteratively. Multiple errors, due to a local concavity of the loop or the presence of multiple loops, might cause the syndrome to grow for some steps, before eventually shrinking once all concavities are removed. In order to be able to fulfill Eq.~\eqref{eq:FlipCommutation} any pair of valid configurations $\mathbf{x}$ and $\mathbf{x}^\prime$ with $\ket{\psi_\mathbf{x}} \neq \ket{\psi_{\mathbf{x}^\prime}}$ must differ by at least three bit-flips.

Let us assume that $\hat{U}$ is a Clifford gate. Then, Pauli strings are mapped to Pauli strings; in particular, for any $\mathbf{i} \in \mathbf{s}_j \cap \text{supp}(\hat{U})$, $\hat{U}\hat{X}_\mathbf{i}\hat{U}^\dagger = \hat{\mathcal{P}}_\mathbf{i}$, where $\hat{\mathcal{P}}_\mathbf{i}$ is a Pauli group operator on $\text{supp}(\hat{U})$. Together with Eqs.~(\ref{eq:UGeneral}, \ref{eq:FlipCommutation}), this further implies for all $\ket{\mathbf{x}}_{\mathbf{s}_j} \in G$
\begin{equation}
    \hat{\mathcal{P}}_\mathbf{i}\ket{\mathbf{x}}_{\mathbf{s}_{j}} \ket{\psi_{\mathbf{x}}}_{\mathbf{s}_{j+1}} = \hat{X}_\mathbf{i}\ket{\mathbf{x}}_{\mathbf{s}_{j}} \ket{\psi_{\mathbf{x}}}_{\mathbf{s}_{j+1}}
    \label{eq:PauliString}
\end{equation}
We now consider two different valid input states $\ket{\mathbf{x}}_{\mathbf{s}_j}$ and $\ket{\mathbf{x}^\prime}_{\mathbf{s}_j} =\prod_\mathbf{i} \hat{X}_\mathbf{i} \ket{\mathbf{x}}_{\mathbf{s}_j}$, where the product runs over some qubits $\mathbf i \in \mathbf s_j$. We find
\begin{align}
    &\hat{U}\ket{\mathbf{x^\prime}}_{\mathbf{s}_{j}}\ket{0 \cdots 0}_{\mathbf{s}_{j+1}} = \hat{U}\left(\prod_\mathbf{i} \hat{X}_\mathbf{i} \right)\ket{\mathbf{x}}_{\mathbf{s}_j}\ket{0 \cdots 0}_{\mathbf{s}_{j+1}} \nonumber \\ &= \left(\prod_\mathbf{i} \hat{\mathcal{P}}_\mathbf{i} \right)\hat{U}\ket{\mathbf{x}}_{\mathbf{s}_j}\ket{0 \cdots 0}_{\mathbf{s}_{j+1}} = \left(\prod_\mathbf{i} \hat{\mathcal{P}}_\mathbf{i} \right)\ket{\mathbf{x}}_{\mathbf{s}_j}\ket{\psi_\mathbf{x}}_{\mathbf{s}_{j+1}} \nonumber \\ &= (-1)^\sigma\left(\prod_\mathbf{i} \hat{X}_\mathbf{i} \right)\ket{\mathbf{x}}_{\mathbf{s}_j}\ket{\psi_\mathbf{x}}_{\mathbf{s}_{j+1}} = (-1)^\sigma\ket{\mathbf{x}^\prime}_{\mathbf{s}_j}\ket{\psi_\mathbf{x}}_{\mathbf{s}_{j+1}}.
\end{align}
From the first to the second line we have used the Clifford property of $\hat{U}$. From the second to the third line we have used Eq.~(\ref{eq:PauliString}) and the commutation relation $\hat{\mathcal{P}}_{\mathbf{j}}\hat{X}_{\mathbf{i}} = \pm \hat{X}_{\mathbf{i}}\hat{\mathcal{P}}_{\mathbf{j}}$, as both are Pauli group operators. Consecutively applying the Pauli group operators $\hat{\mathcal{P}}_{\mathbf{i}}$ gives rise to the phase factor $(-1)^\sigma$, as the arising $\hat X_{\mathbf{i}}$ operators have to be pulled through the remaining Pauli strings $\hat{\mathcal{P}}_{\mathbf{j}\neq\mathbf{i}}$. The argument is the same for $\hat{Y}$ operators.

Comparing this with Eq.~(\ref{eq:UGeneral}), this can hold true only if $\ket{\psi_{\mathbf{x}^\prime}}_{\mathbf{s}_{j+1}} = \pm \ket{\psi_{\mathbf{x}}}_{\mathbf{s}_{j+1}}$. As the input states $\ket{\mathbf{x}}_{\mathbf{s}_{j}}$ and $\ket{\mathbf{x}^\prime}_{\mathbf{s}_{j}}$ were arbitrary, it follows that if $\hat{U}$ is Clifford, the output state has to be the same for all valid $\ket{\mathbf{x}}_{\mathbf{s}_{j}}$ up to a sign factor, $\ket{\psi_{\mathbf x}}_{\textbf{s}_{j+1}} = \pm \ket{\psi}_{\mathbf s_{j+1}} \forall \mathbf x$, where $\ket{\psi}_{\mathbf s_{j+1}}$ is some state. The error-free state can thus not be entangled between adjacent slices $\mathbf{s}_j$ and $\mathbf{s}_{j+1}$, as for any superposition of input states $\hat U \left(\sum_{\mathbf x}  \alpha_{\mathbf x}\ket{\mathbf x}_{\mathbf s_j}\right)\ket{0\cdots 0}_{\mathbf s_{j+1}} = \left(\sum_{\mathbf x}  (-1)^{\sigma_{\mathbf x}}\alpha_{\mathbf x}\ket{\mathbf x}_{\mathbf s_j}\right)\ket{\psi}_{\mathbf s_{j+1}}$, where $\sigma_{\mathbf x} = 0,1$ encode the signs.
In summary, if a gate $\hat{U}$ of the form in Eq.~(\ref{eq:UGeneral}) is both stable against Pauli noise and Clifford, the targeted state can be neither long-range ordered nor long-range entangled.

\let\oldaddcontentsline\addcontentsline
\renewcommand{\addcontentsline}[3]{}
\bibliography{sources.bib}
\let\addcontentsline\oldaddcontentsline

\newpage
\leavevmode \newpage

\setcounter{equation}{0}
\setcounter{page}{1}
\setcounter{figure}{0}
\renewcommand{\thepage}{S\arabic{page}}  
\renewcommand{\thefigure}{S\arabic{figure}}
\renewcommand{\theequation}{S\arabic{equation}}
\onecolumngrid
\begin{center}
\textbf{\large{Supplemental Material:}}\\
\textbf{\large{\papertitle}}\\ \vspace{10pt}
\end{center}

\maketitle

\twocolumngrid

\section{A. Relation between unitary circuits and isometric Tensor Network States}
In the main text, we briefly mention that the unitary circuits we propose for generating CSS states can be related to their tensor network representations. In this section, we elaborate on this connection. We consider a tensor network state (TNS) representation on a $D$ dimensional lattice system which has open boundary conditions on one boundary, which hosts some boundary state $\ket{\Psi_b}$; the other boundaries can be open or periodic. In each unit cell, a tensor $T^\mathbf{p}_{\mathbf v}$ is placed with a set of $N_p$ physical indices $\mathbf p = (p_1,\dots, p_{N_p})$, which represent the physical degrees of freedom in this unit cell (for simplicity, we assume them to be qubits, i.e., $p_j = 0,1$), and a set of $N_v$ virtual indices $\mathbf v = (v_1, \dots, v_{N_v})$, each of which connects the tensor to some neighboring tensor. In general, the graph spanned by the virtual legs between tensors can differ from the actual lattice. The TNS wavefunction is obtained by contracting the virtual legs,
\begin{equation}
        \ket{\Psi} = \sum_{\mathbf p_1 \cdots \mathbf p_k} \text{Tr}\left( \left\{ T^{\mathbf p_1}, \cdots, T^{\mathbf p_k} \right\} \right) \ket{\mathbf p_1\cdots\mathbf p_k},
    \label{eq:TensorWF}
\end{equation}
where the sum runs over physical configurations in all unit cells (enumerated from $1$ to $k$) and $\text{Tr}$ denotes the tensor contraction over the virtual legs with boundary conditions imposed by the state $\ket {\Psi_b}$.

The fixed point states we consider are equal-weight superpositions of $\hat Z$ basis state configurations; this allows for a much simpler form of the tensors $T^\mathbf{p}_{\mathbf v}$. We split up the virtual legs $\mathbf v$ into two sets which we label outgoing legs $\mathbf o = (o_1, \dots, o_{N_o})$ and incoming legs $\mathbf i =(i_1, \dots, i_{N_i})$, respectively. Each outgoing leg $o_j \in \mathbf o$ is locked with some physical leg $p_j \in \mathbf p$; that means, the entry of the tensor is 0 if $o_j \neq p_j$. Multiple outgoing legs are allowed to correspond to the same physical leg. Thus, the outgoing legs feed forward information about the physical configuration in this unit cell to the tensors they connect to. Likewise, the incoming legs $\mathbf i$ provide information about the physical configuration in the unit cells they connect to. Thus, the information contained in the tensor $T$ is stored entirely in the legs $\mathbf i$ and $\mathbf p$,
\begin{equation}
    T_{\mathbf{i}, \mathbf{o}}^{\mathbf{p}} =  \left(\prod_{o_j \in \mathbf{o}}\delta_{o_j, p_j} \right)W_{\mathbf{i}, \mathbf{p}},
    \label{eq:WDecomposition}
\end{equation}
where $\delta_{a,b}$ is the Kronecker delta. Here, we have introduced the matrix $W$, whose columns are labeled by the configurations of the incoming legs $\mathbf{i}$ and its rows labeled by configurations of the physical legs $\mathbf{p}$~\cite{liu_simulating2024, boesl_quantum2025, boesl2025skeleton}.

As the states are equal-weight superpositions, every non-zero entry of $W$ has the same magnitude. There is a subset $G$ of ``valid" incoming configurations $\mathbf{i}$ which allow all local stabilizer conditions to be fulfilled; $W$ can be rescaled so that the corresponding columns are square normalized, $\sum_{\mathbf p} \vert W_{(\mathbf i, \mathbf p)} \vert^2 =1$ for all $\mathbf i \in G$. All other incoming configurations necessarily imply some stabilizer to be violated, the corresponding column thus has to be empty $\sum_{\mathbf p} \vert W_{(\mathbf i, \mathbf p)} \vert^2 =0$ for all $\mathbf i \notin G$. We can now introduce a modified matrix $W^\prime$ which agrees with the original $W$ in all columns labeled by valid incoming configurations, $W^\prime_{(\mathbf i, \mathbf p)} = W_{(\mathbf i, \mathbf p)}$ for all $\mathbf i \in G$, but with additional non-zero entries in the columns corresponding to the invalid configurations such that the entire matrix is square normalized,
\begin{equation}
    \sum_{\mathbf p} \vert W^\prime_{(\mathbf i, \mathbf p)} \vert^2 =1 \; \forall \mathbf i.
    \label{eq:Normalization}
\end{equation}
So far, we have not specified how these additional entries are chosen. If the boundary state $\ket{\Psi_b}$ respects all stabilizers, the wavefunction $\ket{\Psi^\prime}$ obtained from the tensor contraction of the tensors $(T^\prime)_{\mathbf{i}, \mathbf{o}}^{\mathbf{p}}$ obtained from $W^\prime$ via Eq.~(\ref{eq:WDecomposition}) necessarily agrees with $\ket{\Psi}$ as no invalid incoming configuration is ever encountered, i.e., we can orient the tensor network so that $\ket{\Psi_b}$ corresponds to exclusively incoming legs.

The condition (\ref{eq:Normalization}) is strongly reminiscent of an isometric TNS (isoTNS)~\cite{zaletel_isometric_2020}. For these TNS, the local tensors are linear isometric maps from a subset of virtual legs to the physical legs and the remaining virtual legs. Indeed, the TNS discussed here can be made into an isoTNS by first adding in each unit cell additional physical degrees of freedom which simply copy the $\hat Z$ basis states, $\ket {\mathbf p} \rightarrow \bigotimes^N_k \ket{\mathbf p}_k$, and then re-assorting these physical degrees of freedom to adjacent tensors. It should be noted that in isoTNS parlance, the notions of ``incoming" and ``outgoing" legs are exchanged~\cite{zaletel_isometric_2020}.

The isoTNS property guarantees the existence of a sequential unitary circuit which generates the quantum state from an initial state $\ket{\Psi_b}\ket{0\cdots 0}$~\cite{slattery_quantum_2021, wei_sequential_2022}; the depth of this circuit scales with linear system size $L$. Likewise, a matrix respecting condition (\ref{eq:Normalization}) implies a local unitary $\hat U$ exists which can be used to sequentially prepare the state in linear depth. Such a unitary takes the qubits corresponding to the virtual legs $\mathbf i$ in adjacent unit cells as control input and acts on the physical degrees of freedom $\mathbf p$, so that the following identity holds
\begin{align}
    &\hat U \ket{{i_1} \cdots {i_{N_i}}}\ket{0 \cdots 0} =\nonumber\\& \sum_{\substack{(p_1, \dots, p_{N_p}) \\\in \{0,1\}^{N_p}}} W^\prime_{(\mathbf i, \mathbf p)} \ket{{i_1} \cdots {i_{N_i}}}\ket{p_1\cdots p_{N_p}}.
\end{align}
Eq.~(\ref{eq:Normalization}) guarantees that the state $\ket{{i_1} \cdots {i_{N_i}}}\ket{0 \cdots 0}$ is mapped to a normalized state. The action on the states orthogonal to all $\ket{{i_1} \cdots {i_{N_i}}}\ket{0 \cdots 0}$ can in principle be chosen arbitrarily; of course, certain prescriptions lead to simpler circuit constructions of the associated unitary $\hat U$.

We can use the aforementioned freedom of choice in defining $W^\prime$ to optimize the circuit stability against bit-flip noise. If the subspace $G$ is the entire set of possible incoming configurations $\mathbf i$, i.e., $W^\prime$ always agrees with $W$, there is no way to detect whether a bit-flip error has occurred on these qubits before applying $\hat U$. This is the case for example for the 1D GHZ state or for the 2D toric code. If $G$ is not the entire set of configurations of $\mathbf i$, the unitary $\hat U$ can detect whether an error has happened on these qubits as the associated configuration can never satisfy all stabilizers. If it is possible to uniquely identify the most probably valid configuration, i.e., the one closest under bit-flips, the associated gate $\hat U$ can effectively ignore a single error and prevent its proliferation. In practice, these are the majority-vote/curvature-shrinking based procedures introduced in the main text for symmetry breaking order in 3D or the 4D (3,1) toric code. If multiple valid configurations are closest to such erroneous configurations, it cannot be locally decided which choice of $W^\prime$ or $\hat U$ counteracts the error most efficiently; however, via stabilizer measurements, a classical decoder can be used to find the right set of unitaries for each unit cell in the current slice, again preventing error proliferations. Examples for this are the 2D GHZ state, the 3D toric code and the 4D (2,2) toric code.

We illustrate this construction with the example of the GHZ state $\ket{\text{GHZ}} = \frac{1}{\sqrt 2} (\ket{0 \cdots 0} + \ket{1 \cdots 1})$, the qubits living on the vertices of a cubic lattice in three dimensions. The tensor $T^p_{\mathbf i, \mathbf o}$ describing this state has a single physical degree of freedom representing the qubit and three incoming $\mathbf i = (i_1, i_2, i_3)$ and outgoing $\mathbf o = (o_1, o_2, o_3)$ legs, one in each lattice direction. The corresponding matrix $W_{\text{GHZ}}$ is
\begin{equation}
    \begin{blockarray}{ccccccccc}
         \vert 000 \rangle & \vert 001 \rangle & \vert 010 \rangle & \vert 011 \rangle & \vert 100 \rangle & \vert 101 \rangle & \vert 110 \rangle & \vert 111 \rangle & \\
        \begin{block}{(cccccccc)c}
            1 & 0 & 0 & 0 & 0 & 0 & 0 & 0 & \;\vert 0 \rangle \\
             0 & 0 & 0 & 0 & 0 & 0 & 0 & 1 &\; \vert 1 \rangle \\
        \end{block}
    \end{blockarray}\, .
\end{equation}
All columns of incoming configurations where the $i$ indices do not agree are 0. Different ways of filling them yield different unitary circuits. For example, a simple rule would be to have an entry be non-zero if the physical degree of freedom $p$ agrees with one incoming leg, e.g. $i_1$. The modified matrix $W^\prime_1$ is
\begin{equation}
    \begin{blockarray}{ccccccccc}
         \vert 000 \rangle & \vert 001 \rangle & \vert 010 \rangle & \vert 011 \rangle & \vert 100 \rangle & \vert 101 \rangle & \vert 110 \rangle & \vert 111 \rangle & \\
        \begin{block}{(cccccccc)c}
            1 & 1 & 1 & 1 & 0 & 0 & 0 & 0 & \;\vert 0 \rangle \\
             0 & 0 & 0 & 0 & 1 & 1 & 1 & 1 &\; \vert 1 \rangle \\
        \end{block}
    \end{blockarray}\, ,
\end{equation}
and the corresponding local unitary gate is a bit-flip conditioned on the qubit $i_1$, $\hat U_1 = \text{CNOT}_{i_1, p}$. However, the generation circuit built from this unitary $\hat U_1$ is unstable to bit-flip noise as it consists simply of independent repetition codes on single qubits. 

Alternatively, we can choose that entry in a column to be non-zero which agrees with the majority of $\mathbf i$ legs; then, the modified matrix $W^\prime_2$ is 
\begin{equation}
    \begin{blockarray}{ccccccccc}
         \vert 000 \rangle & \vert 001 \rangle & \vert 010 \rangle & \vert 011 \rangle & \vert 100 \rangle & \vert 101 \rangle & \vert 110 \rangle & \vert 111 \rangle & \\
        \begin{block}{(cccccccc)c}
            1 & 1 & 1 & 0 & 1 & 0 & 0 & 0 & \;\vert 0 \rangle \\
             0 & 0 & 0 & 1 & 0 & 1 & 1 & 1 &\; \vert 1 \rangle \\
        \end{block}
    \end{blockarray}\, ,
\end{equation}
and the unitary from this matrix is $\hat U_{\text{SSB}}$ given in Eq.~(\ref{eq:MajorityRule}) in the main text. Crucially, all three $W$ matrices given above yield the same state $\ket{\text{GHZ}}$ upon contraction provided the boundary state $\ket {\Psi_b}$ is in a GHZ superposition. However, only the circuit implementing the gates $\hat U_{\text{SSB}}$ is stable to bit-flip noise below a threshold value $p_c$ as shown in the main text.

\section{B. Errors on qubits outside of the control slice}
In the main text, we have only considered Pauli errors on qubits in the slice $\mathbf{s}_j$ before the unitary $\hat{U}_{\mathbf{s}_j, \mathbf{s}_{j+1}}$ is applied. In this section, we briefly argue why errors on other qubits during this time step do not pose a conceptual problem, as they either map to local physical noise or are eventually treated by our error-correcting gates.

We may first consider an error on a qubit in an earlier slice $\mathbf{s}_k$ with $k < j$. By construction, this qubit has already been fully prepared, i.e., there is no unitary that acts on it in this or later time steps. Accordingly, this merely constitutes local Pauli noise, against which the considered long-range correlated or topologically ordered states are robust up to a finite error threshold.

Secondly, an error may happen on a qubit in a later slice $\mathbf{s}_k$ with $k > j$, which has not been acted on so far, including the to-be prepared slice $\mathbf{s}_{j+1}$. As the qubits in this region are all initialized in the $\ket{0}$ state, $\hat{Z}$ operators can be neglected as they act trivially on this state. $\hat X$ operators on the other hand stay local Pauli noise: As each layer $\hat U_{\mathbf s_j, \mathbf s_{j+1}}$ is constructed from commuting local unitaries and each qubit $\mathbf i \in \mathbf s_{j+1}$ in the second layer is acted on only by one local unitary $U$, we need to consider only the effect of this local unitary, $\hat{U}_{\mathbf{s}_j, \mathbf{s}_{j+1}} \hat X_{\mathbf{i}}\hat{U}_{\mathbf{s}_j, \mathbf{s}_{j+1}}^\dagger = \hat U \hat X_{\mathbf i} \hat U^\dagger$. We do this explicitly for the considered examples. For the circuit generating the long-range ordered state $\hat \rho_{\text{SSB}}$, the $\hat X$ error commutes with the gate $\hat U_{\text {SSB}}$, Eq.~(\ref{eq:MajorityRule}), $\hat U_{\text{SSB}} \hat X_{\mathbf i} \hat U_{\text{SSB}}^\dagger = \hat X_{\mathbf i}$, and the effect of an $\hat X$ before application of the gate is the same as if it were applied after, where it is treated by the next layer unitary $\hat{U}_{\mathbf{s}_{j+1}, \mathbf{s}_{j+2}}$ which takes $\mathbf{s}_{j+1}$ as input. Similarly, for the unitary $\hat U_{(3,1)\text{TC}}$, Eq.~(\ref{eq:TCUnitaryAppMod}), an $\hat X$ error on the first qubit $\mu_1$ is equivalent to a $\hat Z$ error after application of the gate, $\hat U_{(3,1)\text{TC}} \hat X_{\mu_1} \hat U_{(3,1)\text{TC}}^\dagger = \hat Z_{\mu_1}$, while on the other three qubits $\mu_2, \mu_3, \mu_4$ it commutes with the gate, $\hat U_{(3,1)\text{TC}} \hat X_{\mu_i} \hat U_{(3,1)\text{TC}}^\dagger = \hat X_{\mu_i}$. Both are thus treated by the unitary $\hat{U}_{\mathbf{s}_{j+1}, \mathbf{s}_{j+2}}$.  Finally, $\hat{Y} = i\hat X \hat Z$ errors can be viewed as a combination of the former two and thus are treated likewise.

\section{C. Geometry and Action of the 4D Toric Code Unitary}
In this section, we discuss the geometry and action of the local unitary $\hat U _{(3,1)\text{TC}}$ used to generate the 4D $(3,1)$ toric code. As explained in the main text, the layers $\hat{U}_{\mathbf{s}_j, \mathbf{s}_{j+1}}$ of the sequential generation circuit act between adjacent 3D slices of the hypercubic lattice which are orthogonal to the fully diagonal $w$ direction; i.e., this is the $(1,1,1,1)$ direction in terms of the lattice vectors $\mathbf{e}_i$, $i =1,\dots,4$ of the lattice.

Each layer is decomposed into local gates $\hat{U}_{(3,1)\text{TC}}$, each of which is supported on 16 qubits. It takes 12 qubits $\sigma_1, \dots, \sigma_{12}$ in the first layer $\mathbf{s}_j$ as control input in the configurational $\hat{Z}$ basis and acts on the four qubits $\mu_1, \dots, \mu_4$ in the second layer $\mathbf{s}_{j+1}$ conditioned on the configuration of the $\sigma$ qubits. Each $\sigma$ qubit in the layer $\mathbf{s}_{j}$ serves as a control qubit in four distinct local gates. The total layer $\hat{U}_{\mathbf{s}_j, \mathbf{s}_{j+1}}$ can thus be decomposed into four layers of $\hat{U}_{(3,1)\text{TC}}$ gates, where in each of these layers the individual local gates do not overlap and can thus be applied simultaneously.

\begin{figure}
    \centering
    \def\c(#1,#2,#3,#4){
        ({(#1+#2-2*#3-#4)/2}, {(#1-#2-2*#3+#4)/2}, {(-#1+#2-2*#3+#4)/2})
    }

\def\cc(#1,#2,#3,#4){
        ({(#1+#2-2*#3-#4)/2}, {(#1-#2-2*#3+#4)/2-1}, {(-#1+#2-2*#3+#4)/2+1})
    }
\begin{tikzpicture}[scale=3, >=Stealth, lattice3dv2, dot/.style={circle, fill=black, minimum size=4pt, inner sep=0pt}, dot_big/.style={circle, draw, color = black,fill=white, minimum size=4pt, inner sep=0pt}, plus/.style={
        circle, draw, fill=white, minimum size=4pt, inner sep=0pt, thick,
        path picture={
            \draw[thick] (path picture bounding box.north) -- (path picture bounding box.south);
            \draw[thick] (path picture bounding box.east) -- (path picture bounding box.west);
        }
    }
    ]

    \begin{scope}[shift={(0,0)}]


    \draw[->, thick, black, line width = 1 pt] \cc(0, 0, 0, 0) -- \cc(0, -0.4, 0, 0);
    \draw[->, thick, black, line width = 1 pt] \cc(0, 0, 0, 0) -- \cc(-0.4, 0, 0, 0);
    \draw[->, thick, black, line width = 1 pt] \cc(0, 0, 0, 0) -- \cc(0, 0, -1.0, 0);
    \draw[->, thick, black, line width = 1 pt] \cc(0, 0, 0, 0) -- \cc(0, 0, 0, -0.4);
    
    \node[black] at \cc(-0.5, 0, 0, 0) {$\mathbf{e}_1$};
    \node[black] at \cc(0, -0.5, 0, 0) {$\mathbf{e}_2$};
    \node[black] at \cc(0, 0, -1.2, 0) {$\mathbf{e}_3$};
    \node[black] at \cc(0, 0, 0, -0.5) {$\mathbf{e}_4$};

    \fill[red!20, opacity = 0.5] \c(0.00, 0.00, 0.00, 0.00) -- \c(1.10, 0.00, 0.00, 0.00) -- \c(1.10, 1.10, 0.00, 0.00) -- \c(0.00, 1.10, 0.00, 0.00) -- cycle;
    \fill[red!20, opacity = 0.5] \c(0.00, 0.00, 0.00, 0.00) -- \c(1.10, 0.00, 0.00, 0.00) -- \c(1.10, 0.00, 1.10, 0.00) -- \c(0.00, 0.00, 1.10, 0.00) -- cycle;
    \fill[red!20, opacity = 0.5] \c(0.00, 0.00, 0.00, 0.00) -- \c(1.10, 0.00, 0.00, 0.00) -- \c(1.10, 0.00, 0.00, 1.10) -- \c(0.00, 0.00, 0.00, 1.10) -- cycle;
    \fill[red!20, opacity = 0.5] \c(0.00, 0.00, 0.00, 0.00) -- \c(0.00, 1.10, 0.00, 0.00) -- \c(0.00, 1.10, 1.10, 0.00) -- \c(0.00, 0.00, 1.10, 0.00) -- cycle;
    \fill[red!20, opacity = 0.5] \c(0.00, 0.00, 0.00, 0.00) -- \c(0.00, 1.10, 0.00, 0.00) -- \c(0.00, 1.10, 0.00, 1.10) -- \c(0.00, 0.00, 0.00, 1.10) -- cycle;
    \fill[red!20, opacity = 0.5] \c(0.00, 0.00, 0.00, 0.00) -- \c(0.00, 0.00, 1.10, 0.00) -- \c(0.00, 0.00, 1.10, 1.10) -- \c(0.00, 0.00, 0.00, 1.10) -- cycle;

    \draw[thick, gray!60,line width = 1 pt] \c(0, -0.15, 0, 0) -- \c(0, 1.15, 0, 0);
    \draw[thick, gray!60,line width = 1 pt] \c(-0.15, 0, 0, 0) -- \c(1.15, 0, 0, 0);
    \draw[thick, gray!60,line width = 1 pt] \c(0, 0, -0.15, 0) -- \c(0, 0, 1.15, 0);
    \draw[thick, gray!60,line width = 1 pt] \c(0, 0, 0, -0.15) -- \c(0, 0, 0, 1.15);

    \draw[thick, gray!40,line width = 1 pt, dashed] \c(1, 0, 0, 0) -- \c(1, 0, 0, 1.15);
    \draw[thick, gray!40,line width = 1 pt, dashed] \c(1.15, 0, 0, 1) -- \c(0, 0, 0, 1);
    \draw[thick, gray!40,line width = 1 pt, dashed] \c(-0.15, 1, 0, 0) -- \c(1.15, 1, 0, 0);
    \draw[thick, gray!40,line width = 1 pt, dashed] \c(-0.15, 0, 1, 0) -- \c(1.15, 0, 1, 0);
    
    \draw[thick, gray!40,line width = 1 pt, dashed] \c(0, 1, 0, 0) -- \c(0, 1, 0, 1.15);
    \draw[thick, gray!40,line width = 1 pt, dashed] \c(0, 1.15, 0, 1) -- \c(0, 0, 0, 1);
    \draw[thick, gray!40,line width = 1 pt, dashed] \c(1, -0.15, 0, 0) -- \c(1, 1.15, 0, 0);
    \draw[thick, gray!40,line width = 1 pt, dashed] \c(0, -0.15, 1, 0) -- \c(0, 1.15, 1, 0);
    
    \draw[thick, gray!40,line width = 1 pt, dashed] \c(0, 0, 1, 0) -- \c(0, 0, 1, 1.15);
    \draw[thick, gray!40,line width = 1 pt, dashed] \c(0, 0, 1.15, 1) -- \c(0, 0, 0, 1);
    \draw[thick, gray!40,line width = 1 pt, dashed] \c(1, 0, -0.15, 0) -- \c(1, 0, 1.15, 0);
    \draw[thick, gray!40,line width = 1 pt, dashed] \c(0, 1, -0.15, 0) -- \c(0, 1, 1.15, 0);
    
    \node[dot_big] (o1) at \c(0.5, 0, 0, 0) {};
    \node[dot_big] (o2) at \c(0, 0.5, 0, 0) {};
    \node[dot_big] (o3) at \c(0, 0, 0.5, 0) {};
    \node[dot_big] (o4) at \c(0, 0, 0, 0.5) {};
    \node[right] (o1) at \c(0.5, 0, 0, 0) {$\mu_1$};
    \node[above] (o2) at \c(0, 0.5, 0, 0) {$\mu_2$};
    \node[left] (o3) at \c(0, 0, 0.5, 0) {$\mu_3$};
    \node[above] (o4) at \c(0, 0, 0, 0.5) {$\mu_4$};


    \node[dot] (i1) at \c(1, 0.5, 0, 0) {};
    \node[dot] (i2) at \c(0.5, 1, 0, 0) {};
    \node[dot] (i3) at \c(1, 0, 0.5, 0) {};
    \node[dot] (i4) at \c(0.5, 0, 1, 0) {};
    \node[dot] (i5) at \c(1, 0, 0, 0.5) {};
    \node[dot] (i5) at \c(0.5, 0, 0, 1) {};
    \node[dot] (i5) at \c(0, 1, 0.5, 0) {};
    \node[dot] (i5) at \c(0, 0.5, 1, 0) {};
    \node[dot] (i5) at \c(0, 1, 0, 0.5) {};
    \node[dot] (i5) at \c(0, 0.5, 0, 1) {};
    \node[dot] (i5) at \c(0, 0, 1, 0.5) {};
    \node[dot] (i5) at \c(0, 0, 0.5, 1) {};

    \node[above] (i1) at \c(1, 0.5, 0, 0) {$\sigma_1$};
    \node[left] (i2) at \c(0.5, 1, 0, 0) {$\sigma_2$};
    \node[right] (i3) at \c(1, 0, 0.5, 0) {$\sigma_3$};
    \node[left] (i4) at \c(0.5, 0, 1, 0) {$\sigma_4$};
    \node[above] (i5) at \c(1, 0, 0, 0.5) {$\sigma_5$};
    \node[right] (i5) at \c(0.5, 0, 0, 1) {$\sigma_6$};
    \node[left] (i5) at \c(0, 1, 0.5, 0) {$\sigma_7$};
    \node[above] (i5) at \c(0, 0.5, 1, 0) {$\sigma_8$};
    \node[above] (i5) at \c(0, 1, 0, 0.5) {$\sigma_9$};
    \node[above] (i5) at \c(0, 0.5, 0, 1) {$\sigma_{10}$};
    \node[above] (i5) at \c(0, 0, 1, 0.5) {$\sigma_{11}$};
    \node[right] (i5) at \c(0, 0, 0.5, 1) {$\sigma_{12}$};


    
    \node[red] at \c(0, 0.7, 0, 0.7) {\large$\hat{U}_{(3,1)\text{TC}}$};
    
    \end{scope}
\end{tikzpicture}
    \caption{Support of the $\hat{U}_{(3,1)\text{TC}}$ gate: The qubits $\mu_1, \dots, \mu_4$ (white dots) span a unit cell in slice $\mathbf{s}_{n+1}$ and are prepared conditioned on the qubits $\sigma_1, \dots, \sigma_{12}$ (black dots) in slice $\mathbf{s}_n$. The gate prepares the $\mu$ qubits in such a state as to minimize the violations of the $\hat Z$ stabilizers $\hat A_p$ on the six plaquettes $p$ in the support of $\hat{U}_{(3,1)\text{TC}}$, each of which is shared by two $\sigma$ and two $\mu$ qubits. The total sequential circuit acts in the $w$ direction, which is fully diagonal relative to all lattice directions $\mathbf{e}_i$ (top left corner).}
    \label{fig:TC_gate_action}
\end{figure}
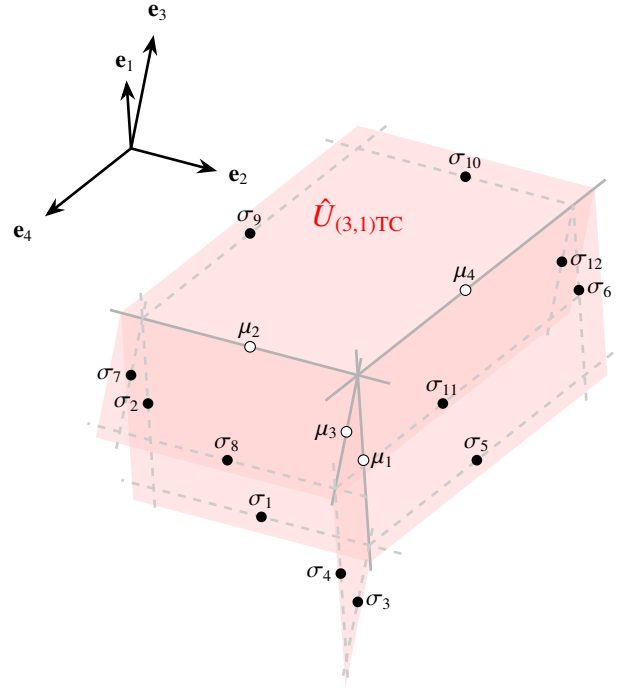

The precise geometry of one local gate is visualized in Fig.~\ref{fig:TC_gate_action}. The four $\mu$ qubits live on four edges which share an adjacent vertex $v$, i.e., they constitute a unit cell of the 4D hypercubic lattice. Each pair of these four qubits $\mu_i, \mu_j$ shares a plaquette $p_{ij}$ with two distinct $\sigma$ qubits. In total, there are thus six plaquettes in the support of one unitary $\hat{U}_{(3,1)\text{TC}}$.

Formally, the unitary is defined as
\begin{equation}
\begin{aligned}
    &\hat{U}_{(3,1)\text{TC}} \ket{\sigma_1\cdots \sigma_{12}}\ket{\mu_1 \cdots \mu_4} = \\& \left(\prod_{i \in F_{\sigma_1\cdots\sigma_{12}}} \hat X_{\mu_i} \right) \left(\prod_{i=2}^4 \text{CNOT}_{\mu_1, \mu_i}\right)\hat H_{\mu_1} \ket{\sigma_1\cdots \sigma_{12}}   \ket{\mu_1\cdots\mu_4},
    \label{eq:TCUnitaryAppMod}
\end{aligned}
\end{equation}
where $\hat H$ is the Hadamard gate $\hat H \ket{0/1} = (\ket 0 \pm \ket 1)/\sqrt 2$, and $F_{\sigma_1\cdots\sigma_{12}}$ is a subset of $\{2,3,4\}$ which we specify in the following for each $\sigma$ configuration. This gate can be decomposed into two steps: First, a Clifford operation on the $\mu$ qubits consisting of the Hadamard gate and three CNOT gates; the relevant action of this gate is  $\left(\prod_{i=2}^4 \text{CNOT}_{1, i}\right)\hat H_{1} \ket{0000} = (\ket{0000} + \ket{1111})/\sqrt 2$. Then, as a second step, we apply a set of bit-flips conditioned on the $\sigma$ qubits. This is chosen so that, assuming that the $\mu$ qubits are in the state $\ket{\mu_1\cdots \mu_4} = \ket{0000}$, $\hat U_{(3,1)\text{TC}}$ prepares them in an equal-weight superposition of two $\hat{Z}$ basis states for which the smallest number of plaquette stabilizers $\hat A_{p} = \prod_{\mathbf{i} \in \partial p} \hat{Z}_{\mathbf{i}}$ in the support of the gate are violated, leading to Eq.~(\ref{eq:TCUnitary}). The gate thus performs a majority vote on all implicated $\hat{A}_{p}$ stabilizers. The amplitudes are equal to further ensure that the vertex stabilizers $\hat B_v = \prod_{\mathbf i \in \delta v} \hat X_{\mathbf i}$ are also respected.

In slightly different terms, the gate ensures that as many plaquettes $p_{ij}$ as possible have even parity $\pi_{ij} = \left( \sum_{\sigma \in p_{ij}} \sigma\right) \text{mod} \; 2$. For certain ``valid" input configurations, this condition can be fulfilled for all plaquettes at the same time, e.g., $\ket{\sigma_1 \cdots \sigma_{12}} = \ket{0 \cdots 0}$. In this case, there are exactly two valid $\mu$ configurations which are related by a full spin-flip, $\ket{\mu_1\cdots \mu_4}$ and $\prod_{i = 1}^4 X_ i\ket{\mu_1\cdots \mu_4}$. For $\ket{\sigma_1 \cdots \sigma_{12}} = \ket{0 \cdots 0}$, these two options are $\ket{\mu_1 \cdots \mu_4} = \ket{0000}$ and $\ket{\mu_1 \cdots \mu_4} = \ket{1111}$, i.e., $F_{0\cdots 0}$ is the empty set.

Other $\sigma$ configurations will always violate at least one plaquette parity; this indicates that some error has occurred during the preparation. For example, all configurations which differ from a valid configuration by a single bit-flip, like $\ket{\sigma_1 \cdots \sigma_{12}} =  \hat X_{i}\ket{0 \cdots 0}$, fall in this category. For these, the best choice of $\mu$ configurations is the set which is shared by all their closest valid $\sigma$ configurations. Accordingly, the gate prepares the output superposition associated with the nearest valid configuration, i.e., it assumes that only an isolated bit-flip error has occurred, and ignores it.

All other invalid $\sigma$ configurations need two bit-flips to be reached from a valid configuration; for each, there are three distinct classes of valid $\sigma$ configurations with different corresponding $\mu$ configurations. In fact, the different possible pairs of bit-flips which connect them to valid configurations are always the same. We can fix one set of flips (this could be, for example, $\hat X_{\sigma_1}$ and $\hat X_{\sigma_{12}}$, see Fig.~\ref{fig:TC_gate_action}) and assign to each of these invalid $\sigma$ configurations the two $\mu$ configurations of the valid $\sigma$ configuration reached by these bit-flips. This implies there are three equivalent prescriptions for $\hat U_{(3,1)\text{TC}}$. In our cellular automaton numerics, for each local gate we choose this rule randomly in every run. One could also choose a deterministic assignment where each layer $\mathbf s_j$ has a fixed rule for all gates, with the chosen prescription changing periodically between layers.

An isolated bit-flip error on the control input is thus not propagated by the gate $\hat U_{(3,1)\text{TC}}$, see Eq.~(\ref{eq:FlipCommutation}), as it is corrected in one time-step. More generally, this allows for stable preparation at low error density; the majority vote takes into account the local curvature of the violated stabilizer patterns (which form closed loops in the current time slice) and shrinks them considering this curvature.

\begin{figure}
    \centering
    \def\c(#1,#2,#3,#4){
        ({(#1+#2-2*#3-#4)/2}, {(#1-#2-2*#3+#4)/2}, {(-#1+#2-2*#3+#4)/2})
    }
\begin{tikzpicture}[scale=1, >=Stealth, lattice3dv2]

    \begin{scope}[shift={(0,0)}]
    \fill[red!10, opacity=0.7] \c(0,0,0,0) -- \c(0, 1,0, 0) -- \c(0, 1, 1, 0) -- \c(0, 0, 1, 0) -- cycle;
    \fill[red!10, opacity=0.7] \c(0,0,0, 0) -- \c(1, 0,0, 0) -- \c(1, 0, 1, 0) -- \c(0, 0, 1, 0) -- cycle;
    \fill[red!15, opacity=0.8] \c(0,0,0, 0) -- \c(0, 0,0, 1) -- \c(0, 0, 1, 1) -- \c(0, 0, 1, 0) -- cycle;    
    \node[] (x) at \c(1.5, -0.1, 0, 0) {$\mathbf e_1$};
    \node[] (y) at \c(0, 1.45, 0, 0) {$\mathbf e_2$};
    \node[] (z) at \c(0, 0.1, 1.5, 0) {$\mathbf e_3$};
    \node[] (w) at \c(0, 0, 0, 1.45) {$\mathbf e_4$};
    
    \draw[thick, gray!60,line width = 1 pt] \c(0, -0.25, 0, 0) -- \c(0, 1.25, 0, 0);
    \draw[thick, gray!60,line width = 1 pt] \c(-0.25, 0, 0, 0) -- \c(1.25, 0, 0, 0);
    \draw[thick, gray!60,line width = 1 pt] \c(0, 0, -0.25, 0) -- \c(0, 0, 1.25, 0);
    \draw[thick, gray!60,line width = 1 pt] \c(0, 0, 0, -0.25) -- \c(0, 0, 0, 1.25);

    \draw[thick, gray!40,line width = 1 pt, dashed] \c(-0.25, 0, 0, 0) -- \c(1.25, 0, 0, 0);
    \draw[thick, gray!40,line width = 1 pt, dashed] \c(1, 0, 0, 0) -- \c(1, 0, 0, 1.25);
    \draw[thick, gray!40,line width = 1 pt, dashed] \c(1.25, 0, 0, 1) -- \c(0, 0, 0, 1);
    \draw[thick, gray!40,line width = 1 pt, dashed] \c(-0.25, 1, 0, 0) -- \c(1.25, 1, 0, 0);
    \draw[thick, gray!40,line width = 1 pt, dashed] \c(-0.25, 0, 1, 0) -- \c(1.25, 0, 1, 0);
    
    \draw[thick, gray!40,line width = 1 pt, dashed] \c(0, -0.25, 0, 0) -- \c(0, 1.25, 0, 0);
    \draw[thick, gray!40,line width = 1 pt, dashed] \c(0, 1, 0, 0) -- \c(0, 1, 0, 1.25);
    \draw[thick, gray!40,line width = 1 pt, dashed] \c(0, 1.25, 0, 1) -- \c(0, 0, 0, 1);
    \draw[thick, gray!40,line width = 1 pt, dashed] \c(1, -0.25, 0, 0) -- \c(1, 1.25, 0, 0);
    \draw[thick, gray!40,line width = 1 pt, dashed] \c(0, -0.25, 1, 0) -- \c(0, 1.25, 1, 0);
    
    \draw[thick, gray!40,line width = 1 pt, dashed] \c(0, 0, -0.25, 0) -- \c(0, 0, 1.25, 0);
    \draw[thick, gray!40,line width = 1 pt, dashed] \c(0, 0, 1, 0) -- \c(0, 0, 1, 1.25);
    \draw[thick, gray!40,line width = 1 pt, dashed] \c(0, 0, 1.25, 1) -- \c(0, 0, 0, 1);
    \draw[thick, gray!40,line width = 1 pt, dashed] \c(1, 0, -0.25, 0) -- \c(1, 0, 1.25, 0);
    \draw[thick, gray!40,line width = 1 pt, dashed] \c(0, 1, -0.25, 0) -- \c(0, 1, 1.25, 0);
    
    \draw[thick,line width = 1 pt, dashed, red!60!black] \c(-0, 0.5, 0.5, 0) -- \c(0.5, 0.5, 0.5, 0.5);
    \draw[thick,line width = 1 pt, dashed, red!60!black] \c(0.5, -0, 0.5, 0) -- \c(0.5, 0.5, 0.5, 0.5);
    \draw[thick,line width = 1 pt, dashed, red!60!black] \c(0, 0, 0.5, 0.5) -- \c(0.5, 0.5, 0.5, 0.5);

    \draw[thick,line width = 2 pt, red!60!black] \c(0, 0, 0, 0) -- \c(0, 0, 1, 0);
    
    \node[anchor=north] (T) at (0., -0.25, -1.2) {t};
    
    \end{scope}

    \begin{scope}[shift={(0,0, -4)}]
    
    \fill[red!10, opacity=0.7] \c(0,0,0, 0) -- \c(0, 1,0, 0) -- \c(0, 1, 1, 0) -- \c(0, 0, 1, 0) -- cycle;
    \fill[red!15, opacity=0.7] \c(0,0,0, 0) -- \c(0, 1,0, 0) -- \c(0, 1, 0, 1) -- \c(0, 0, 0, 1) -- cycle;
    \fill[red!10, opacity=0.7] \c(0,0,0 ,0) -- \c(1, 0,0, 0) -- \c(1, 0, 1, 0) -- \c(0, 0, 1, 0) -- cycle;
    \fill[red!15, opacity=0.8] \c(0,0,0, 0) -- \c(1, 0, 0, 0) -- \c(1, 0, 0, 1) -- \c(0, 0, 0, 1) -- cycle;
    \node[] (x) at \c(1.5, -0.1, 0, 0) {$\mathbf e_1$};
    \node[] (y) at \c(0, 1.45, 0, 0) {$\mathbf e_2$};
    \node[] (z) at \c(0, 0.1, 1.5, 0) {$\mathbf e_3$};
    \node[] (w) at \c(0, 0, 0, 1.45) {$\mathbf e_4$};
    \draw[thick, gray!60,line width = 1 pt] \c(0, -0.25, 0, 0) -- \c(0, 1.25, 0, 0);
    \draw[thick, gray!60,line width = 1 pt] \c(-0.25, 0, 0, 0) -- \c(1.25, 0, 0, 0);
    \draw[thick, gray!60,line width = 1 pt] \c(0, 0, -0.25, 0) -- \c(0, 0, 1.25, 0);
    \draw[thick, gray!60,line width = 1 pt] \c(0, 0, 0, -0.25) -- \c(0, 0, 0, 1.25);

    \draw[thick, gray!40,line width = 1 pt, dashed] \c(-0.25, 0, 0, 0) -- \c(1.25, 0, 0, 0);
    \draw[thick, gray!40,line width = 1 pt, dashed] \c(1, 0, 0, 0) -- \c(1, 0, 0, 1.25);
    \draw[thick, gray!40,line width = 1 pt, dashed] \c(1.25, 0, 0, 1) -- \c(0, 0, 0, 1);
    \draw[thick, gray!40,line width = 1 pt, dashed] \c(-0.25, 1, 0, 0) -- \c(1.25, 1, 0, 0);
    \draw[thick, gray!40,line width = 1 pt, dashed] \c(-0.25, 0, 1, 0) -- \c(1.25, 0, 1, 0);
    
    \draw[thick, gray!40,line width = 1 pt, dashed] \c(0, -0.25, 0, 0) -- \c(0, 1.25, 0, 0);
    \draw[thick, gray!40,line width = 1 pt, dashed] \c(0, 1, 0, 0) -- \c(0, 1, 0, 1.25);
    \draw[thick, gray!40,line width = 1 pt, dashed] \c(0, 1.25, 0, 1) -- \c(0, 0, 0, 1);
    \draw[thick, gray!40,line width = 1 pt, dashed] \c(1, -0.25, 0, 0) -- \c(1, 1.25, 0, 0);
    \draw[thick, gray!40,line width = 1 pt, dashed] \c(0, -0.25, 1, 0) -- \c(0, 1.25, 1, 0);
    
    \draw[thick, gray!40,line width = 1 pt, dashed] \c(0, 0, -0.25, 0) -- \c(0, 0, 1.25, 0);
    \draw[thick, gray!40,line width = 1 pt, dashed] \c(0, 0, 1, 0) -- \c(0, 0, 1, 1.25);
    \draw[thick, gray!40,line width = 1 pt, dashed] \c(0, 0, 1.25, 1) -- \c(0, 0, 0, 1);
    \draw[thick, gray!40,line width = 1 pt, dashed] \c(1, 0, -0.25, 0) -- \c(1, 0, 1.25, 0);
    \draw[thick, gray!40,line width = 1 pt, dashed] \c(0, 1, -0.25, 0) -- \c(0, 1, 1.25, 0);

    \draw[thick,line width = 1 pt, dashed, red!60!black] \c(0, 0.5, 0.5, 0) -- \c(0.5, 0.5, 0.5, 0.5);
    \draw[thick,line width = 1 pt, dashed, red!60!black] \c(0.5, 0, 0.5, 0) -- \c(0.5, 0.5, 0.5, 0.5);
    \draw[thick,line width = 1 pt, dashed, red!60!black] \c(0, 0.5, 0, 0.5) -- \c(0.5, 0.5, 0.5, 0.5);
    \draw[thick,line width = 1 pt, dashed, red!60!black] \c(0.5, 0, 0, 0.5) -- \c(0.5, 0.5, 0.5, 0.5);
    
    \draw[thick,line width = 2 pt, red!60!black] (0, 0, 0) -- \c(0, 0, 1, 0);
    \draw[thick,line width = 2 pt, red!60!black] (0, 0, 0) -- \c(0, 0, 0, 1);
    \node[anchor=north] (T2) at (0., -0.25, -1.2) {t};
    
    \end{scope}

    \begin{scope}[shift={(0,4, 0)}]
    \draw[thick, gray!60,line width = 1 pt] \c(0, -0.25, 0, 0) -- \c(0, 1.25, 0, 0);
    \draw[thick, gray!60,line width = 1 pt] \c(-0.25, 0, 0, 0) -- \c(1.25, 0, 0, 0);
    \draw[thick, gray!60,line width = 1 pt] \c(0, 0, -0.25, 0) -- \c(0, 0, 1.25, 0);
    \draw[thick, gray!60,line width = 1 pt] \c(0, 0, 0, -0.25) -- \c(0, 0, 0, 1.25);

    \draw[thick, gray!40,line width = 1 pt, dashed] \c(-0.25, 0, 0, 0) -- \c(1.25, 0, 0, 0);
    \draw[thick, gray!40,line width = 1 pt, dashed] \c(1, 0, 0, 0) -- \c(1, 0, 0, 1.25);
    \draw[thick, gray!40,line width = 1 pt, dashed] \c(1.25, 0, 0, 1) -- \c(0, 0, 0, 1);
    \draw[thick, gray!40,line width = 1 pt, dashed] \c(-0.25, 1, 0, 0) -- \c(1.25, 1, 0, 0);
    \draw[thick, gray!40,line width = 1 pt, dashed] \c(-0.25, 0, 1, 0) -- \c(1.25, 0, 1, 0);
    
    \draw[thick, gray!40,line width = 1 pt, dashed] \c(0, -0.25, 0, 0) -- \c(0, 1.25, 0, 0);
    \draw[thick, gray!40,line width = 1 pt, dashed] \c(0, 1, 0, 0) -- \c(0, 1, 0, 1.25);
    \draw[thick, gray!40,line width = 1 pt, dashed] \c(0, 1.25, 0, 1) -- \c(0, 0, 0, 1);
    \draw[thick, gray!40,line width = 1 pt, dashed] \c(1, -0.25, 0, 0) -- \c(1, 1.25, 0, 0);
    \draw[thick, gray!40,line width = 1 pt, dashed] \c(0, -0.25, 1, 0) -- \c(0, 1.25, 1, 0);
    
    \draw[thick, gray!40,line width = 1 pt, dashed] \c(0, 0, -0.25, 0) -- \c(0, 0, 1.25, 0);
    \draw[thick, gray!40,line width = 1 pt, dashed] \c(0, 0, 1, 0) -- \c(0, 0, 1, 1.25);
    \draw[thick, gray!40,line width = 1 pt, dashed] \c(0, 0, 1.25, 1) -- \c(0, 0, 0, 1);
    \draw[thick, gray!40,line width = 1 pt, dashed] \c(1, 0, -0.25, 0) -- \c(1, 0, 1.25, 0);
    \draw[thick, gray!40,line width = 1 pt, dashed] \c(0, 1, -0.25, 0) -- \c(0, 1, 1.25, 0);
    \node[anchor=north] (T1) at (0., 0.6, -1.2) {t+1};
    \node[] (x) at \c(1.5, -0.1, 0, 0) {$\mathbf e_1$};
    \node[] (y) at \c(0, 1.45, 0, 0) {$\mathbf e_2$};
    \node[] (z) at \c(0, 0.1, 1.5, 0) {$\mathbf e_3$};
    \node[] (w) at \c(0, 0, 0, 1.45) {$\mathbf e_4$};
    \end{scope}
    \draw[->, thick] (T.south) to[bend right=30] node[below=2pt] {Single Flip} (T1.south);
    \draw[->, thick, right = 1] (T2.south) to[bend right=30] node[right=7pt] {Double Flip} (T1.south);
\end{tikzpicture}
    \caption{Action of the cellular automaton decoder within a single hyper-cubic unit cell of the 4D $(3, 1)$ toric code, which consists of the qubits on four edges, each pointing in one lattice direction $\mathbf{e}_i$ (full lines). Going from time step $t$ to $t +1$, the decoder checks whether the stabilizers $\hat A_p$ on the six indicated plaquettes are violated and applies $\hat X$ operators based on the result. A single erroneous qubit in the unit cell flips three adjacent stabilizers. Alternatively, two erroneous qubits violate four plaquettes. The local rule of the decoder thus flips two edges if four plaquette stabilizers are violated, and a single edge if its three adjacent stabilizers are violated. This systematically shrinks the syndrome membrane in $(1, 1, 1, 1)$ direction.}
    \label{fig:CA_action_decoding_4D}
\end{figure}
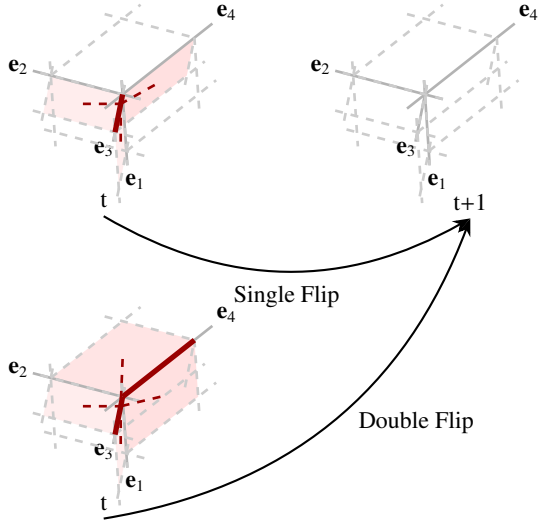

\section{D. Decoding Algorithm for the 4D $(3,1)$ toric code}
In this section, we define the cellular automaton decoder that we use to remove the error syndrome $S(\hat{\rho}_{\text{out}})$ from the noisy output state $\hat{\rho}_{\text{out}}$ which is generated by the 4D toric code sequential circuit in the presence of bit-flip errors, i.e., $\hat X$ noise. Accordingly, $S(\hat{\rho}_{\text{out}})$ contains information about which plaquette stabilizers $\hat A_p$ are violated.

The cellular automaton decoder is adapted from Ref.~\cite{kubica_cellular-automaton_2019}. The decoder is a quantum channel which, via repeated application of a local rule at subsequent time steps $t$, iteratively reduces the syndrome until an error-free state $\hat{\rho}_{(3,1)\text{TC}}$ is reached. This local rule acts on every unit cell in the four-dimensional hypercubic lattice, which is spanned by the qubits living on four edges adjacent to a vertex, each pointing in one lattice direction $\mathbf{e}_i$. The rule takes into account six adjacent plaquettes (these plaquettes are the same as in the construction of the unitary gate $\hat U_{(3,1)\text{TC}}$, see Fig.~\ref{fig:TC_gate_action}) and checks which of the corresponding stabilizers $\hat A_p$ are violated. It then applies $\hat X$ flips according to two sub-rules: Within the unit cell, an edge is shared by exactly three of the indicated plaquettes. The primary rule dictates that an edge be flipped if all three corresponding stabilizers are violated. Due to the high geometric connectivity of the 4D lattice, more complex corners of the syndrome configuration can manifest from two adjacent erroneous edges within a single unit cell. To handle these corners in the syndrome, the secondary rule flips two edges simultaneously if four internal plaquette stabilizers are found to be violated (see Fig.~\ref{fig:CA_action_decoding_4D} for a visualization). For this rule, there is a degree of freedom as such a configuration of four violated stabilizers can be created by two distinct pairs of flipped qubits; we always choose the option including the edge oriented along the $\mathbf{e}_4$-axis.

The action of the decoder can be seen as a generalization of the two-dimensional Toom's rule~\cite{toomsrule}: The violated plaquette stabilizers in $S(\hat{\rho}_{\text{out}})$ appear as closed membranes, i.e., closed two-dimensional manifolds. The decoder iteratively shrinks these manifolds along the fully diagonal $(1,1,1,1)$ direction, i.e., in circuit direction, by locally measuring the curvature of the error syndrome. We assume the boundary state $\ket{\Psi_b}$ to be error-free; even if $\ket{\Psi_b}$ included some errors, the decoder could be extended by a step that removes the error syndrome on this first layer, which would increase the decoding time only by a factor, and does not change the bulk properties of the state.

Cellular automaton decoders will clear arbitrary syndrome configurations in a number of steps which scales at most linearly with linear system size, ${N_{d}} = \mathcal{O}(L)$. The relevant question for our output state $\rho_{\text{out}}$ is whether, for a given error probability $p_{\hat X}$, $N_{d}$ scales sub-linearly in $L$, ${N_{d}} = o(L)$. If this is the case, the decoding channel connecting $\hat \rho_{\text{out}}$ to $\hat \rho_{(3,1)\text{TC}}$ remains local, establishing some kind of non-trivial order (see App.~A). Indeed, as shown in Fig.~\ref{fig:relative_unitary_circuit_depth.pdf}, below a certain threshold probability $p_{\hat X}^c$, the ratio $N_{d}/L$ decreases with increasing $L$, indicating a sublinear relation. 

\section{E. Diameter of Error Syndrome Clusters for the (3,1) Toric Code}
As discussed in the main text and App.~A, to formally establish that the noise output state $\hat \rho_{\text{out}}$ is in the same mixed-state phase as the error-free fixed point state $\hat{\rho}_{(3,1)\text{TC}}$, a local channel mapping one state to the other has to be found for both directions. The decoder presented in the main text serves as a channel mapping $\hat \rho_{\text{out}}$ to $\hat{\rho}_{(3,1)\text{TC}}$. For the opposite direction, we would naturally assume that the physical error model corresponding to Pauli errors in the circuit is local. Below the threshold probability $p_{\hat X}^c$, such a local noise channel should produce only small error clusters. In this section, we provide additional numerical data that the error cluster below $p_{\hat X}^c$ indeed has extent vanishing compared to the system size.

We consider only $\hat X$ errors, as $\hat Z$ errors map trivially to local physical error. In particular, we consider some $\hat A_p$ plaquette stabilizer configuration created by the stochastic circuit mimicking the unitary circuit with interspersed $\hat X$ noise. The defective stabilizers corresponding to propagated $\hat{X}$ errors form two-dimensional membranes on the dual lattice surrounding the physical error region. Under a local noise model below the threshold, errors are expected to occur in small, isolated patches whose spatial extent is vanishingly small relative to the system size. To quantify this, we partition the error syndromes into distinct unconnected clusters of violated plaquettes. 

\begin{figure}
\centering
\includegraphics[width=\linewidth]{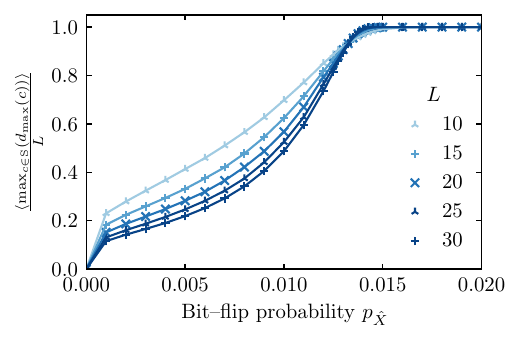}
\caption{Normalized average maximum syndrome cluster diameter $\langle \max_{c \in S} d_{\mathrm{max}}(c) \rangle / L$ as a function of error probability $p_{\hat X}$. Below the threshold $p^c_{\hat X} \approx 0.013$, the relative cluster size decreases towards zero with increasing $L$, demonstrating that errors remain confined to small, isolated patches. Above $p^c_{\hat X}$, the curves cross and saturate at $1$, indicating a transition to a regime dominated by a system-spanning cluster.}
\label{fig:cluster_diameter}
\end{figure}

Two plaquettes $P$ and $Q$ belong to the same cluster $c$ if they are connected through adjacent 3-cells (cubes) $C$, such that $P, Q \in \partial C$. Defined in this way, each cluster is unambiguously isolated from all others. We define the spatial extent of a cluster $c$ as the side lengths $(d_x, d_y, d_z, d_w)$ of the minimal 4D hyper-box fully enclosing $c$. The linear size of the cluster is further defined as the maximum side length, $d_{\mathrm{max}}(c) = \max\{d_x, d_y, d_z, d_w\}$. If a cluster wraps around a boundary along a periodic direction, the side length of its bounding box along that axis is defined to be the linear system size $L$, capping the maximum cluster diameter at $L$.

To evaluate whether the maximal cluster size remains bounded as the lattice expands, we calculate the relative average maximum cluster diameter:\begin{equation}D_{\mathrm{max}} = \frac{\left\langle \max_{c \in S} d_{\mathrm{max}}(c) \right\rangle}{L},\end{equation} where $S$ is the set of clusters in a given error realization in the circuit, and $\langle \cdot \rangle$ denotes the ensemble average over these realizations. We numerically evaluate $D_{\mathrm{max}}$ for system sizes of dimensions $(L, L, L, 2L+1)$, where the directions are the same as in the main text. Periodic boundary conditions are imposed along the $x, y,$ and $z$ directions, while open boundary conditions are applied along the $w$ direction. Due to the open boundaries in $w$, the system contains one fewer layer of complete plaquette stabilizers. Furthermore, the rotated geometry accommodates twice as many unit cells along $w$ for the same spatial length; the chosen dimensions thus ensure that the maximum achievable cluster bounding box remains bounded by $L$.

\begin{figure*}
    \centering
    \includegraphics[]{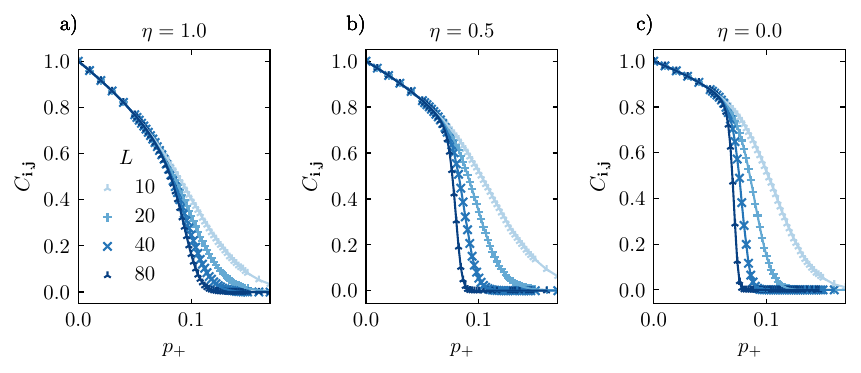}
    \caption{Connected two-site correlations $C_{\mathbf{i}, \mathbf{j}}$ for a sequentially prepared state using the circuit adapting Toom's rule, subject to biased bit-flip noise (Eq.~\eqref{eq:ErrorBiased}) with probabilities $p_+$ and $p_-$, for different values of bias $\eta = \frac{p_-}{p_+}$: a) $\eta=1$, b) $\eta=0.5$, and c) $\eta=0$. For each linear system size $L$, the sites $\mathbf i$ and $\mathbf j$ are separated by a distance of $L/2$ in $z$ direction, which is the direction of the sequential circuit. For small enough probabilities $p_+$, the correlations remain finite, indicating the presence of long-range order, despite the output state $\hat \rho_{\text{out}}$ not being symmetric under bit-flip for $\eta \neq 1$. The transition probability $p_\text{c}$ decreases as the system becomes more biased.}
    \label{fig:toom_corr}
\end{figure*}

As shown in Fig.~\ref{fig:cluster_diameter}, below a critical error rate $p^c_{\hat X} \approx 0.013$, the normalized maximum cluster diameter $D_{\mathrm{max}}$ vanishes asymptotically with increasing system size $L$. This confirms that syndrome defects remain localized and isolated. Above $p^c_{\hat X}$, $D_{\mathrm{max}}$ saturates near unity, signaling a transition where the largest cluster spans the entire linear dimension of the system.

\section{F. Stability against biased noise from Toom's rule}
\label{app:tooms_rule_circuit}
In the main text, we have proposed a simple majority update scheme to prepare a long-range ordered, weakly $\mathbb{Z}_2$ symmetric state in the presence of Pauli noise. However, there are more general noise models that may need a different treatment. One example is biased bit-flip noise acting independently on each qubit $\mathbf{i}$ in the preparation slice $\mathbf{s}_j$, $\mathcal{E}_j = \bigotimes_{\mathbf{i} \in \mathbf{s}_j} \mathcal{E}_\mathbf{i}$. Here, the single qubit channel $\mathcal{E}_\mathbf{i}$ is defined as 
\begin{equation}
 \mathcal{E}_\mathbf{i} (\hat{\rho}) = \hat{E}_{\mathbf{i},0}\hat{\rho}\hat{E}_{\mathbf{i},0}^\dagger + \hat{E}_{\mathbf{i},1}\hat{\rho}\hat{E}_{\mathbf{i},1}^\dagger,
\label{eq:ErrorBiased}
\end{equation}
with the Kraus operators
\begin{align}
    &\hat{E}_{\mathbf{i},0} = \sqrt{1-p_+} \ket 0_{\mathbf i}\bra 0_{\mathbf i} +  \sqrt{1-p_-} \ket 1_{\mathbf i}\bra 1_{\mathbf i}, \nonumber \\     &\hat{E}_{\mathbf{i},1} = \sqrt{p_+} \ket 1_{\mathbf i}\bra 0_{\mathbf i} +  \sqrt{p_-} \ket 0_{\mathbf i}\bra 1_{\mathbf i},
\end{align}
where $0 \leq p_\pm \leq 1$.

The operator $\hat{E}_{\mathbf{i},0}$ represents no error happening, while $\hat{E}_{\mathbf{i},1}$ represents a bit-flip on the state $\ket 0$ with probability $p_+$ and a bit-flip on the state $\ket 1$ with probability $p_-$. The bias $\eta = \frac{p_-}{p_+}$ quantifies how much flipping one $\hat Z$ basis state is preferred over the reverse action.

For $\eta \neq 1$,  $\mathcal{E}_j$ breaks the weak $\mathbb{Z}_2$ symmetry corresponding to a global spin flip $\prod_{\mathbf i} \hat X_{\mathbf i}$ which is respected by standard Pauli noise. The majority vote described in the main text is unstable against this type of noise; in colloquial terms, this rotationally symmetric rule does not shrink an erroneous domain fast enough as any bias $\eta \neq 1$ causes the domain to expand ballistically~\cite{pajouheshgar_exploring2026}. However, there are asymmetric update rules which shrink such domains in linear time and thus provide stability also in the biased case. They can easily be adapted into corresponding quantum circuits; here, we show this for Toom's rule~\cite{toomsrule}, a commonly discussed stable non-equilibrium memory.

Toom's rule acts on bits on a square lattice, and flips a bit $\sigma_{(x, y)} \in \{0, 1\}$ if there is a domain wall north and east of it; it thus shrinks domains in a south-west direction.
This procedure is equivalent to a majority vote on a triple consisting of the spin and its north and east neighbors, $\{\sigma_{(x, y)}, \sigma_{(x+1, y)}, \sigma_{(x, y+1)}\}$. In contrast to the rule performing a majority vote on the spins and all of its four neighbors, this cellular automaton has two steady states of mostly $0$ states and mostly $1$ states, respectively, in the presence of biased bit-flip noise, provided both $p_\pm$ are small.

We can directly translate this rule into a sequential unitary circuit which prepares the state $\hat \rho_{\text{SSB}}$ in the absence of noise. We again consider a cubic lattice of size $L_x \times L_y \times L_z$; here the direction $x,y,z$ correspond to the lattice directions $\mathbf{e}_i$, $i =1,2,3$. The circuit acts in $z$ direction, i.e., each layer unitary $\hat U_{\mathbf s_j, \mathbf s_{j+1}}$ entangles two square layers $\mathbf s_j, \mathbf s_{j+1}$ of size $L_x \times L_y$ which are orthogonal to the $z$ direction. We can use the same local gates $\hat U_{\text{SSB}}$ (Eq.~(\ref{eq:MajorityRule})) as in the main text, simply specifying its support. Each gate applies an $\hat X$ operator on the qubit $\mathbf{i} \in \mathbf s_{j+1}$ with coordinates $\mathbf{i} = (x,y,z = j+1)$, conditioned on the $\hat Z$ basis states of the three qubits $\mathbf j, \mathbf k, \mathbf l \in \mathbf{s}_j$ in the preceding layer with corresponding coordinates $\mathbf j = (x, y, z= j),\, \mathbf k = (x+1, y, z= j), \, \mathbf l = (x, y+1, z= j)$. Each layer unitary $\hat U_{\mathbf s_j, \mathbf s_{j+1}}$ can be decomposed into eight layers of non-overlapping gates $\hat U_\text{SSB}$.

The resultant state $\hat\rho_{\text{out}}$ will still have long-range order in the presence of the biased noise of Eq.~(\ref{eq:ErrorBiased}), provided both $p_\pm$ are below threshold. It should however be noted that for $\eta \neq 1$, $\hat\rho_{\text{out}}$ will not be symmetric under the weak $\mathbb{Z}_2$ symmetry associated to a global spin flip $\prod_\mathbf i \hat X_\mathbf i$ of $\hat \rho_{\text{SSB}}$, as this is explicitly broken by the noise; by definition, it thus cannot be an SSB state. As in the main text, we evaluate the two-point correlations $C_{\mathbf{i} , \mathbf{j} } = \langle \hat{Z}_{\mathbf{i} } \hat{Z}_{\mathbf{j} }\rangle - \langle \hat{Z}_{\mathbf{i} }\rangle \langle \hat{Z}_{\mathbf{j} }\rangle$ by means of cellular automaton simulations. We show the results of the bulk correlations between the two sites $\mathbf i$ and $\mathbf j$ separated by $L/2$ in $z$ direction for different system sizes $L$ as a function of $p_+$ in Fig.~\ref{fig:toom_corr}, considering different bias values $\eta$. As in the majority rule, the correlations remain finite in the thermodynamic limit below a critical probability $p_c$. This transition probability decreases as the bias becomes stronger.

\end{document}